\documentclass[useAMS,usenatbib]{mn2e}

\usepackage{color}
\usepackage{graphicx}
\usepackage{epstopdf}
\usepackage{aas_macros}
\usepackage{times}
\usepackage{amsmath,amsfonts,amssymb}

\title[Discreteness-Driven Relaxation in Cosmological Simulations]
      {Spurious Small-Scale Structure \& Discreteness-Driven Relaxation in Cosmological Simulations}
      \author[Power et al.]
      {C. Power$^{1}$\thanks{chris.power@icrar.org}, A. S. G. Robotham$^1$,
        D. Obreschkow$^1$, A. Hobbs$^2$ \& G. F. Lewis$^3$\\
        $^1$International Centre for Radio Astronomy Research, 
        University of Western Australia, 35 Stirling Highway, Crawley, 
        Western Australia 6009, Australia\\
        $^2$Institute for Astronomy, ETH Zurich, Wolfgang-Pauli-Strasse 27, CH-8093 Zurich, Switzerland\\
        $^3$Sydney Institute for Astronomy, School of Physics,
        A28, The University of Sydney, New South Wales 2006, Australia}
\begin{document}
  
  \date{}
  
  \pagerange{\pageref{firstpage}--\pageref{lastpage}} \pubyear{}
  
  \maketitle
  
  \label{firstpage}
  
  \begin{abstract}
    There is strong evidence that cosmological $N$-body simulations 
    dominated by Warm Dark Matter (WDM) contain spurious or unphysical
    haloes, most readily apparent as regularly spaced low-mass haloes
    strung along filaments. We show that spurious haloes are a
    feature of traditional $N$-body simulations of cosmological
    structure formation models, including WDM and Cold Dark Matter (CDM)
    models, in which gravitational collapse proceeds in an initially
    anisotropic fashion, and arises naturally as a consequence of
    discreteness-driven relaxation. We demonstrate this using controlled
    $N$-body simulations of plane-symmetric collapse and show
    that spurious haloes are seeded at shell crossing by localised velocity
    perturbations induced by the discrete nature of the density field,
    and that their characteristic separation should be approximately the
    mean inter-particle separation of the $N$-body simulation, which is
    fixed by the mass resolution within the volume. Using cosmological
    $N$-body simulations in which particles are split into two collisionless
    components of fixed mass ratio, we find that the spatial distribution of
    the two components show signatures of discreteness-driven relaxation in
    their spatial distribution on both large and small scales. Adopting a
    spline kernel gravitational softening that is of order the comoving mean
    inter-particle separation helps to suppress the effect of
    discreteness-driven relaxation, but cannot eliminate it completely. These
    results provide further motivation for recent developments of new
    algorithms, which include, for example, revisions of the traditional
    $N$-body approach by means of spatially adaptive anistropric gravitational
    softenings or solve explicitly for the evolution of dark matter in phase
    space.
  \end{abstract}
  
  
 \section{Introduction}
 \label{sec:intro}
Cosmological $N$-body simulations are a well established tool for studying 
the formation and non-linear evolution of structure in the Universe. Much 
of what we know about the Cold Dark Matter (CDM) model, the currently favoured
theoretical framework within which we investigate the growth of this structure,
derives from such simulations \citep[e.g.][]{springel.etal.2006}, and they
have revealed that dark matter haloes have central densities that are 
divergent \citep[e.g.][]{navarro.etal.2010}; that the abundance of these
haloes increases with decreasing halo mass $M$ as $M^{-\alpha}$ with 
$\alpha \simeq 1.8$ (see \citealt{murray.etal.2013} for a survey of published 
halo mass functions); and that haloes contain a wealth of substructure, 
remnants of the merging hierarchy by which the halo assembled 
\citep[e.g.][]{springel.etal.2008}, independent of their mass 
\citep[e.g.][]{ishiyama.etal.2013}. 

It can be argued that it is the abundance of small-scale structure -- low-mass
haloes and substructure haloes (hereafter subhaloes) -- that 
is the defining prediction of the CDM model \citep[e.g.][]{power.2013}, and so 
it is crucial that we understand its properties if we are to devise robust
observational tests of the model. This can be appreciated by noting that
alternatives to the CDM model modify its predictions on small scales. For
example, self-interacting dark matter (SIDM) has a finite interaction cross
section whose influence is greatest at high dark matter densities
\citep[e.g.][]{loeb.weiner.2011}, while Warm Dark Matter (hereafter WDM)
free-streams in the early Universe to erase density perturbations that would
otherwise collapse gravitationally to form low-mass haloes
\citep[e.g.][]{bode.etal.2001}.

\begin{figure*}
  \centerline{\includegraphics[width=0.95\textwidth]{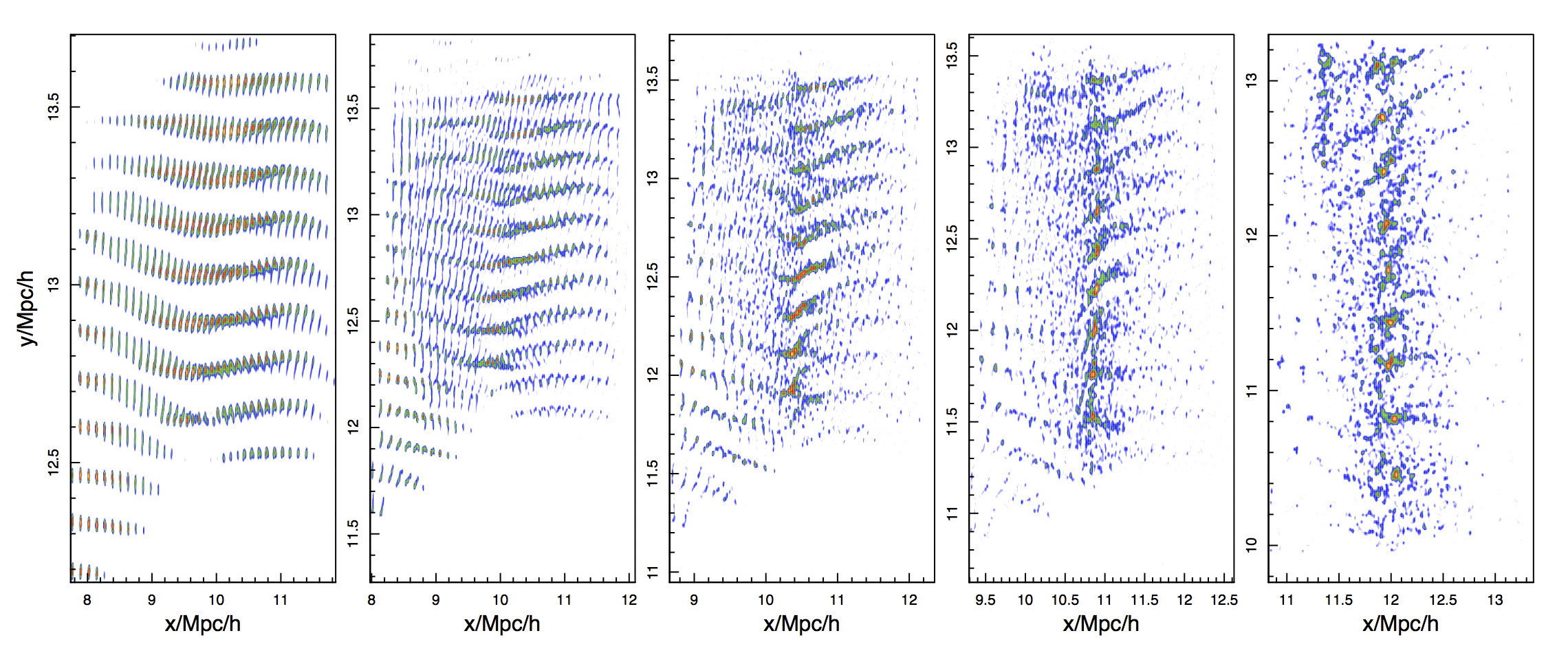}}
  \caption{{\bf ``Beads-on-a-String'' in simulations of the WDM model.} 
    \emph{From left to right:} In the rightmost panel we show an $x$-$y$
    projection of beads-on-a-string, identified in a friends-of-friends
    catalogue at $z$=0, in a
    $N$-body simulation of a 0.5 keV/$c^2$ WDM model. $N$-body particles
    are smoothed using a binned 2D kernel density estimate and coloured
    according to projected density. We track this material from $z$=0 back
    to $z$=20 (leftmost panel) to show that individual knots at late times
    grew from sheets at early times. This kind of
    procedure has been used to `clean' WDM simulations of beads
    \citep[e.g.][]{lovell.etal.2013}. }
  \label{fig:beads_on_a_string}
\end{figure*}

However, modelling structure formation in
alternative dark matter models, especially those such as WDM in which initial
small-scale density perturbations are suppressed, has proven challenging with
the traditional $N$-body approach. Early WDM simulations
\citep[e.g.][]{bode.etal.2001,knebe.etal.2002} appeared to confirm physical
intuition that the abundance of low-mass haloes and subhaloes is suppressed
relative to that in corresponding CDM simulations, but they also revealed the
presence of `beads-on-a-string', regularly spaced low-mass haloes along
filaments (see rightmost panel of Figure~\ref{fig:beads_on_a_string}).
Initially it was argued that these haloes are the low-mass objects we might
expect to form via fragmentation \citep[e.g.][]{knebe.etal.2003}, but
subsequent work demonstrates that they are a numerical artifact
(notably \citealt{wang.white.2007}, but see also the more recent work of
\citealt{myers.etal.2015}, \citealt{sousbie.colombi.2015},
\citealt{hahn.angulo.2015}), with properties
that are sensitive to the mass resolution of the $N$-body
simulation -- the inter-`bead' spacing decreases with decreasing particle mass
as $m_p^{1/3}$, and the halo mass function rises rapidly at low masses
below a mass scale that decreases with increasing mass resolution, again as
$m_p^{1/3}$ \citep[cf.][]{wang.white.2007}. 

\medskip

That such numerical artifacts arise is not surprising, as noted by
e.g. \citet{angulo.2013a}, \citet{hahn.2013a}, \citet{myers.etal.2015},
\citet{sousbie.colombi.2015}, and \citet{hahn.angulo.2015}. Traditional
$N$-body simulations discretise
the cosmic matter density field into particles, which, as noted by
\citet{binney.2004}, introduces localised perturbations into the gravitational
force a particle experiences along its trajectory over time. This implies that
$N$-body particles are subject to velocity perturbations of magnitude,
\begin{equation}
  \label{eq:deltav}
  \delta v \simeq \frac{Gm_p}{\epsilon\,v};
\end{equation}
here $G$ is the gravitational constant, $v$ is the typical peculiar velocity,
which is fixed by the mass distribution, and $\epsilon$ is the $N$-body
particle's gravitational softening length, which is usually chosen to be a
fraction of the typical inter-particle separation within the simulation volume.
The magnitude of these perturbations should remain small relative to the
mean-field velocities induced by the large scale matter distribution
if the evolution of the system is to be treated as collisionless.

In the case of the CDM model, velocity perturbations induced by discretisation
are likely to be difficult to disentangle from velocities of a physical origin
induced by the small-scale density perturbations encoded in the matter power
spectrum. In contrast, these small-scale density perturbations are absent in
WDM models below a particular mass scale, and gravitational collapse of lower
mass objects is either suppressed or delayed
relative to the CDM model; velocity perturbations induced by discretisation
may be larger than the mean-field gravitational peculiar velocity, and so WDM
models and generic dark matter models with reduced initial small-scale density
perturbations should show evidence of these velocity perturbations, which
trigger gravitational collapse and seed the formation of spurious haloes. We
refer to the influence
of these discreteness-induced velocity perturbations on the evolution of
$N$-body particle trajectories as discreteness-driven relaxation.

Figure~\ref{fig:beads_on_a_string} shows `beads-on-a-string' identified
at $z$=0 in a WDM $N$-body simulation that are likely are the product of
discreteness-driven relaxation -- they can be traced back to a Lagrangian
region in the initial conditions that is planar,
distinctly different from the regions that collapse to form physical haloes.
This suggests that spurious haloes could be tagged by inspection of the initial
conditions, and it has motivated efforts to `clean' WDM simulations of
spurious haloes \citep[e.g.][]{lovell.etal.2013,schneider.2013}. However,
spurious haloes are likely to be a generic problem in cosmological
$N$-body simulations, for the reasons outlined above, and so cleaning in this
fashion will only pick out the most obvious instances \citep[a similar point
  is made in][]{hahn.angulo.2015}. Indeed, as \citet{ludlow.porciani.2011}
have noted, identifying the progenitors of low-mass haloes in the linear CDM
density field remains an unsolved problem, which suggests strongly that a
fraction of low-mass CDM haloes are possibly spurious, broken and scattered
beads-on-a-string.

A number of studies have explored how discreteness effects influences the
accuracy of $N$-body simulations \citep[e.g.][]{melott.etal.1997,splinter.etal.1998,knebe.etal.2000,power.etal.2003,binney.knebe.2002,diemand.etal.2004,heitmann.etal.2005,romeo.etal.2008}, as well as hydrodynamical/$N$-body simulations
\citep[e.g.][]{angulo.2013a}; here the focus is on, in general, measurements of the
power spectrum, correlation function, and the internal structure of dark
matter haloes. This study focuses on the role discreteness plays in the
formation of `beads-on-a-string' that are so evident in WDM simulations; we
demonstrate that discreteness-driven relaxation will arise naturally when a
discretised density field is used to model the initially anisotropic phase of
gravitational collapse \citep[e.g.][]{zeldovich.1970,kuhlman.etal.1996} that
is characteristic of realistic cosmological models; we highlight that it affects
CDM simulations, although less readily apparent than in WDM simulations; and
we show that traditional $N$-body methods can suppress discreteness-driven
relaxation through appropriate choice of gravitational softening, but cannot
fully eliminate it. As such, this study provides additional motivation for
innovative new extensions to $N$-body algorithms that have been developed, such
as those that solve for the evolution of dark matter by tracking phase space
elements \citep[e.g.][]{hahn.2013a,hahn.angulo.2015} or by incorporating
gravitational softening that is both spatially adaptive and anisotropic
\citep[cf.][]{hobbs.2015}.

\medskip
In the remainder of this paper, we undertake a series of numerical experiments,
described in \S\ref{sec:sims}, to highlight the influence of
discreteness-driven relaxation in $N$-body simulations of cosmological
structure formation. To develop insights into how anistropic
gravitational collapse might seed the formation of spurious haloes, we simulate
plane-symmetric collapse \citep[cf.][]{zeldovich.1970,shandarin.zeldovich.1989}
in \S\ref{ssec:pancake}, which approximates how the initial collapse proceeds,
and follow its spatial and phase space structure, varying the mass resolution
($m_p$) and the gravitational softening ($\epsilon$), both of
which influence the magnitude of velocity perturbations (Eq~\ref{eq:deltav}),
to identify when and where relaxation becomes important. Applying these insights
to cosmological $N$-body simulations, in \S\ref{ssec:cosmosims}
we consider CDM and WDM models in which
the collisionless component is split into two components; this approach is
inspired by the numerical experiments of \citet{binney.knebe.2002}, who looked
for the signature of two-body relaxation via mass segregation on the internal
structure of dark matter haloes. We investigate the degree to which the choice
of gravitational softening ($\epsilon$) can help to reduce discreteness-driven
relaxation by examining the spatial structure of the dark matter and halo
density fields, the mass functions of haloes, and the effects of mass segregation
within haloes. Finally, in \S\ref{sec:summary} we summarise our results,
assessing the extent to which
cosmological $N$-body simulations are impacted by discreteness-driven
relaxation, and highligthing why novel extensions to the $N$-body are needed.

\section{The Simulations}
\label{sec:sims}

\paragraph*{Plane-Symmetric Collapse Simulations} have an exact solution up to
shell crossing \citep[cf. ][]{zeldovich.1970,shandarin.zeldovich.1989} and
provide a convenient approximation to the initially anisotropic phase of
gravitational collapse that occurs in realistic dark matter models (i.e. CDM
or models in which the CDM power spectrum is suppressed on small spatial
scales). These runs will demonstrate that beads-on-a-string arise
naturally when the gravitational softening $\epsilon$ is smaller than the mean
inter-particle separation $\bar{d}$, and that beading will occur on the
scale of the mean inter-particle separation, which scales with the particle
mass as $\bar{d} = (m_p/\Omega_0\,\bar{\rho})^{1/3} = (V/N)^{1/3}$, where
$\Omega_0$ is the matter density parameter, $\bar{\rho}$ is the mean density,
$V$ is the simulation volume, and $N$ is the number of particles in the
simulation volume.

For our simulation set-up, we follow \citet{hahn.2013a} in adopting an
Einstein de Sitter cosmology
($\Omega_0$=1, $\Omega_{\Lambda}$=0) with a dimensionless Hubble parameter of
$h$=0.7, a periodic box of side $L_{\rm box}=10 h^{-1} \rm Mpc$ and a starting
redshift of $z_{\rm start}=99$; we performed runs with $64^3$, $128^3$ and $256^3$ particles.
The uniform particle distribution is a cubic
mesh and we apply a one-dimensional sinusoidal potential perturbation
$\phi(\vec{x})$ along the $x$-axis, where
\begin{equation}
\phi(\vec{x}) = \phi_0\cos(k_xx)
\end{equation}
where $\vec{x}=(x,y,z)$ and $k_x=2\pi/L_{\rm box}$. We choose the amplitude of
the potential perturbation $\phi_0$ such that shell crossing occurs at $z$=4.
Particle positions and velocities are initialised by applying the
\citet{zeldovich.1970} approximation. In a subset of the runs, we split the particles
into two components with varying mass ratios; see below for further details.

\paragraph*{Two Component Cosmological $N$-body Simulations} allow us to
assess the effects of discreteness-driven relaxation by looking at differences
in the spatial distribution of two collisionless components of fixed mass
ratio that have co-evolved in the same gravitational potential. To do this,
we consider a CDM model and a WDM counterpart (we have runs with
$m_{\rm WDM}=(0.2,0.5) {\rm keV}/c^2$ but concentrate on the $0.2 {\rm keV}/c^2$ runs
to simplify our analysis)
using boxes of side $L_{\rm box}=20 h^{-1} \rm Mpc$ with starting redshift
$z_{\rm start}=99$ and assuming cosmological parameters of $\Omega_0=0.27$, 
$\Omega_{\Lambda}=0.73$, $h=0.705$ and $\sigma_8=0.81$ at $z=0$
\citep[][]{komatsu.etal.2011}. For our two collisionless components, we
consider mass ratios of $1/\sqrt{2}$ and $1/4$ respectively. These runs were
initialised from a regular cubic mesh and the lighter 
components was offset by $\bar{d}/2$ in each of the three dimensions. The fiducial runs
consist of $2\times\,256^3$ particles, but we also have lower resolution runs of
$2\times\,64^3$ particles for comparison with the results of \citet{binney.knebe.2002}.

Initial conditions were created using standard techniques
\citep[e.g.][]{power.etal.2003} -- a statistical realization of a Gaussian 
random density field is generated in Fourier space, with variance given 
by the linear matter power spectrum, and the Zel'dovich approximation is used
to compute initial particle positions and velocities. The power spectrum
for the CDM model is obtained by convolving the primordial power spectrum 
$P(k) \propto k^{n_{\rm spec}}$ with the transfer function appropriate for our 
chosen set of cosmological parameters, computed using the Boltzmann code 
{\texttt{CAMB}} \citep[cf.][]{lewis.etal.2000}. Following 
\citet{bode.etal.2001}, we obtain the initial power spectra for our WDM models 
by filtering the CDM power spectrum with an additional transfer function of 
the form
\begin{equation}
  \label{eq:transfer}
  T^{\rm WDM}(k) = \left(\frac{P^{\rm WDM}(k)}{P^{\rm CDM}(k)}\right)^{1/2}=\left[1+(\alpha\,k)^{2\nu}\right]^{-5/\nu}
\end{equation}
where $k$ is the wave-number; $\nu$=1.2 is a numerical constant; and 
$\alpha$ is a function of the WDM particle mass
\citep[see equation A9 of][]{bode.etal.2001}, which we write as
\begin{equation}
  \label{eq:alpha}
  \alpha=0.0413\left(\frac{\Omega_{\rm X}}{0.27}\right)^{0.15}\left(\frac{h}{0.705}\right)^{1.3}\left(\frac{m_{\rm X}}{{\rm keV}/c^2}\right)^{-1.15}\left(\frac{g_X}{1.5}\right)^{0.29}.
\end{equation}
Here it is assumed that the WDM particle is the thermal relic $X$ with mass
$m_X$; $\Omega_X$ is the global matter density parameter of $X$; and $g_X$ is
the number of spin degrees of freedom, assumed to be 1.5 for WDM
\citep[cf.][]{bode.etal.2001}. For reference, we follow convention and define
a half-mode length $\lambda^{\rm half}$, as the point at which the WDM transfer
function drops to 1/2 \citep[see, e.g.,][]{schneider.etal.2012}; in this case,
\begin{equation}
  \label{eq:half_mode}
  k^{\rm half}=\frac{2\pi}{\lambda^{\rm half}}=\frac{1}{\alpha}\left(2^{\mu/5}-1\right)^{1/2\mu}
\end{equation}
and so the equivalent half-mode mass is,
\begin{equation}
  \label{eq:half_mode_mass}
  M^{\rm half}_{\rm WDM}=\frac{4\pi}{3}\bar{\rho}\left(\frac{\lambda^{\rm half}}{2}\right)^3=\frac{4\pi}{3}\bar{\rho}\left[\pi\alpha\left(2^{\mu/5}-1\right)^{-1/2\mu}\right]^3.
\end{equation}
The corresponding values for the $m_{\rm WDM}$=(0.2,0.5) keV/$c^2$ models are
$M^{\rm half}_{\rm WDM}\simeq (149.8,6.3)\times 10^{10} h^{-1} {\rm M_{\odot}}$.

Note that we do not include an additional velocity to mimic the effects of
free-streaming in the early Universe\footnote{In practice this is done by
  assigning a random velocity component (typically
  drawn from a Fermi-Dirac distribution) to particles in addition to their
  velocities predicted by linear theory
  \citep[cf.][]{klypin.etal.1993,colin.2008,maccio.etal.2012}}; this
is possibly an important omission for the 0.2 keV/$c^2$ run, less so
for the 0.5 keV/$c^2$ run \citep[e.g.][]{colin.2008,angulo.2013}, but
we note that modelling this
effect correctly in a $N$-body simulation is difficult -- it can lead to an
unphysical excess of small-scale power in the initial conditions if the
simulation is started too early (see Figure 1 of
\citealt{colin.2008} for a nice illustration of this problem) -- so for
clarity we ignore this effect \citep[see also discussion in][]{power.2013}.

\paragraph*{Simulation Parameters} All simulations were run using the parallel
TreePM code {\small GADGET2} \citep{springel.2005}. We use the spline kernel
gravitational softening with a comoving softening scale $\epsilon$
\citep[cf. Eq 4 of][]{springel.2005}, such that the density distribution of
a single particle is a convolution of a Dirac $\delta$-function and a normalised kernel and
is expressed as $\tilde{\delta}({\bf{x}})=W(|{\bf{x}}|,2.8\epsilon)$; here
the spline kernel $W(r)$ is written as 
\begin{align}
  W(r,h)=\frac{8}{\pi\,h^3}\left\{
  \begin{array}{ll}
    1-6\left(r/h\right)^2+6\left(r/h\right)^3, & 0 \leq r/h \leq \frac{1}{2} \\
    2\left(1-r/h\right)^3, & \frac{1}{2} \leq r/h \leq 1 \\
    0, & r/h>1. \\
  \end{array}
  \right.
\end{align}
We keep the softening scale $\epsilon$ fixed in comoving coordinates, and use
individual and adaptive timesteps for each particle,
$\Delta t = \eta \sqrt{\epsilon/a}$, where $a$ is the magnitude of a
particle's gravitational acceleration and $\eta=0.05$ determines the accuracy
of the time integration. We used GADGET in standard TreePM mode with a PM dimension
of 512. For our reference softening, we define
$\epsilon_0$=$\bar{d}$ and we considered values of $\epsilon$=(0.01,0.1,1,10)
$\epsilon_0$. We considered cases in which the softening was fixed in
physical units and in which the time integration accuracy was made more
stringent, but these did not affect our general findings. 
For completeness, we have tested the sensitivity of our results to the
underlying TreePM algorithm by running versions of our simulations with
gravitational forces computed both in purely tree mode (i.e. with the
{\small -DPMGRID} flag switched off when compiling {\small GADGET2} and in
TreePM mode (i.e. with {\small -DPMGRID} switched on) with a PM grid dimension
that varies between 64 and 1024 by factors of 2. Full details of these
tests are presented in the appendix.
\medskip
  
\paragraph*{Halo Identification} We use the {\small SubFind} algorithm of 
\citet{springel.etal.2001} to generate friends-of-friends (FOF) catalogues,
assuming a linking length of $b$=0.2$\bar{d}$. For each FOF group we determine
its centre-of-density 
$\vec{r}_{\rm cen}$ using the iterative ``shrinking spheres'' algorithm 
and identify this as the halo centre \citep[cf.][]{power.etal.2003}. From 
this, we calculate quantities such as virial radius $r_{\rm vir}$, which we
define as the radius at which the mean interior density is
$\Delta_{\rm vir}=200$ times the critical density of the Universe at that
redshift, $\rho_{\rm c}(z)=3H^2(z)/8\pi G$, where $H(z)$ and $G$ are the Hubble
parameter at $z$ and the gravitational constant respectively. The corresponding 
virial mass $M_{\rm vir}$ is 
\begin{equation}
  \label{eq:mvir}
        {M_{\rm vir}=\frac{4\pi}{3} \Delta_{\rm vir} \rho_{\rm c} r_{\rm vir}^3.}
\end{equation}

\section{Results}
\label{sec:results}

\subsection{Plane-Symmetric Collapse}
\label{ssec:pancake}

\begin{figure*}
  \centerline{\includegraphics[width=5.3cm]{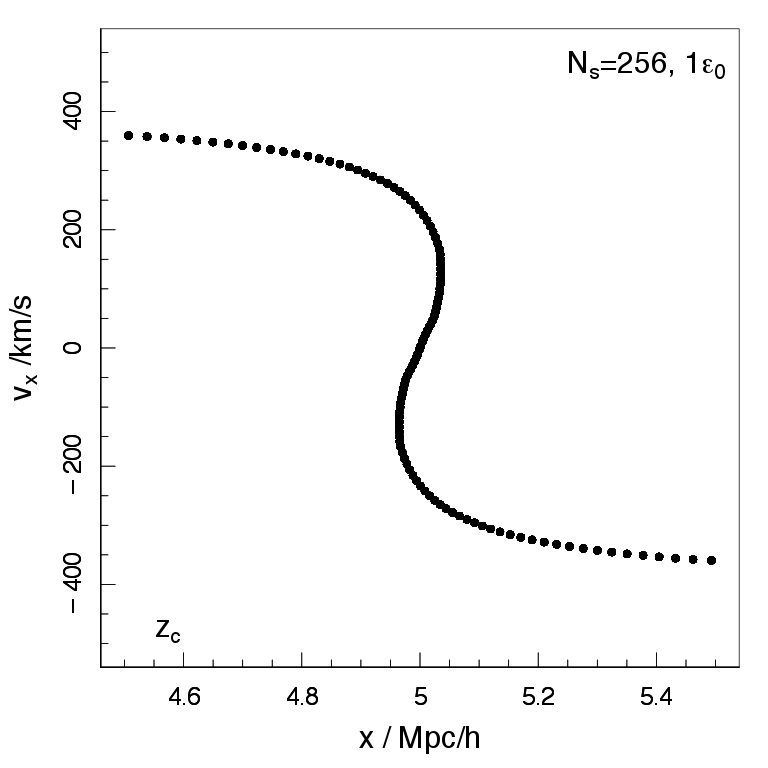}
    \includegraphics[width=5.3cm]{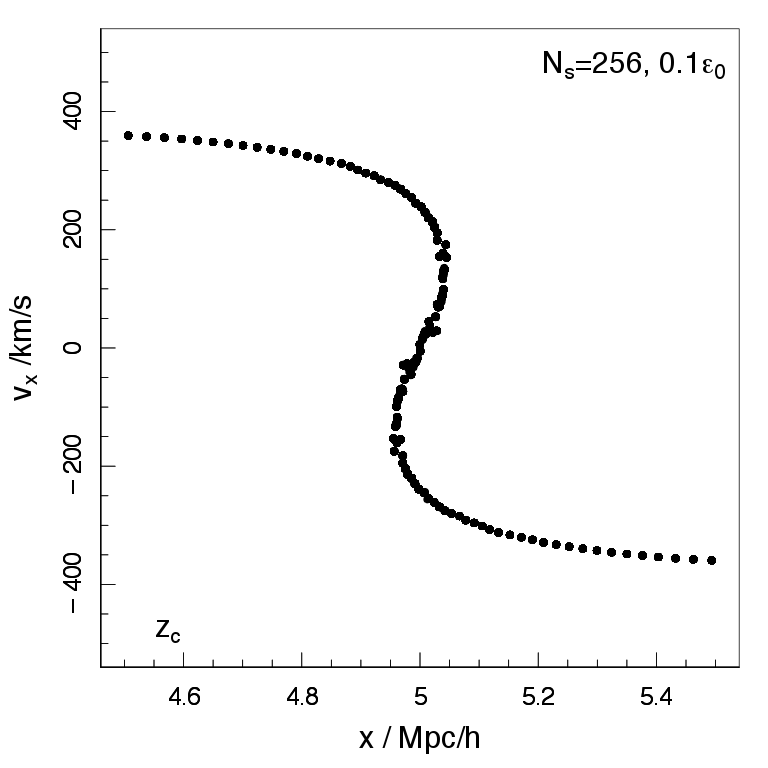}
    \includegraphics[width=5.3cm]{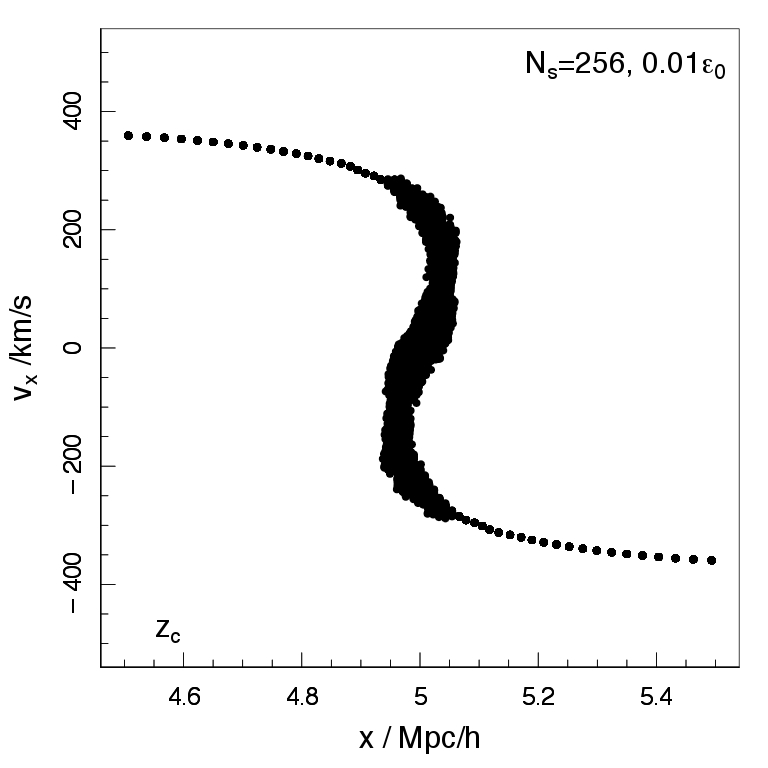}}
  \centerline{\includegraphics[width=5.3cm]{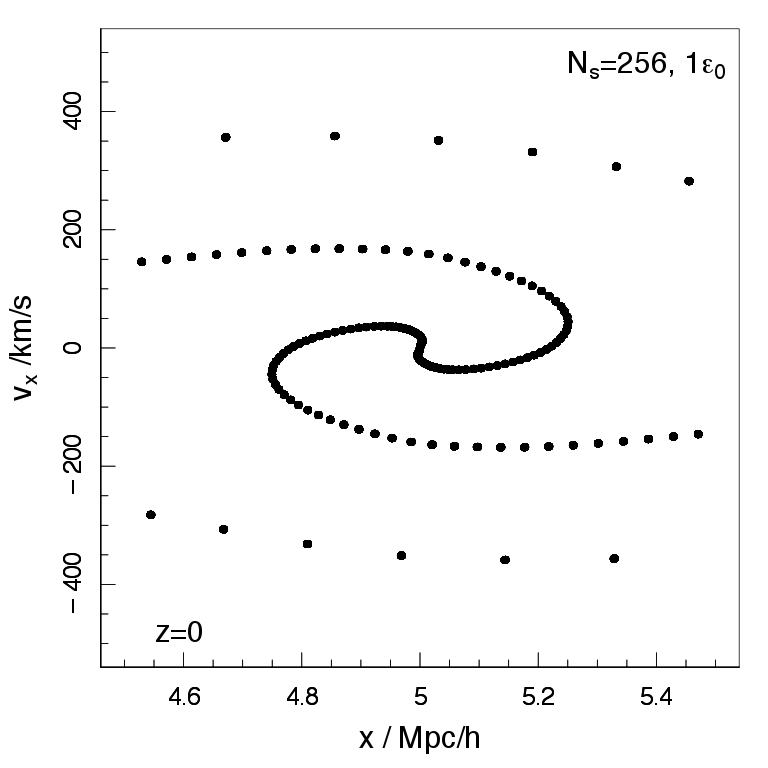}
    \includegraphics[width=5.3cm]{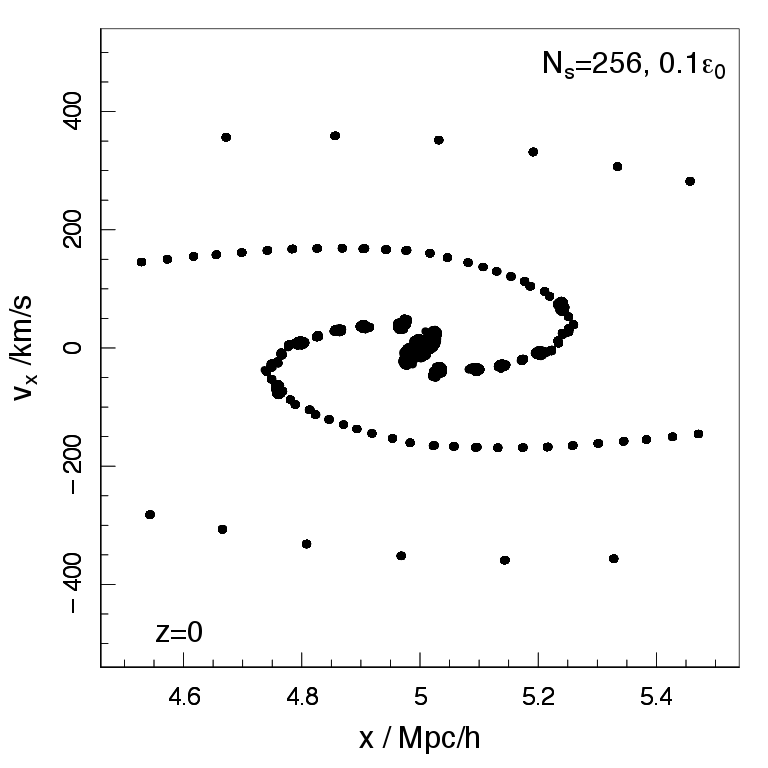}
    \includegraphics[width=5.3cm]{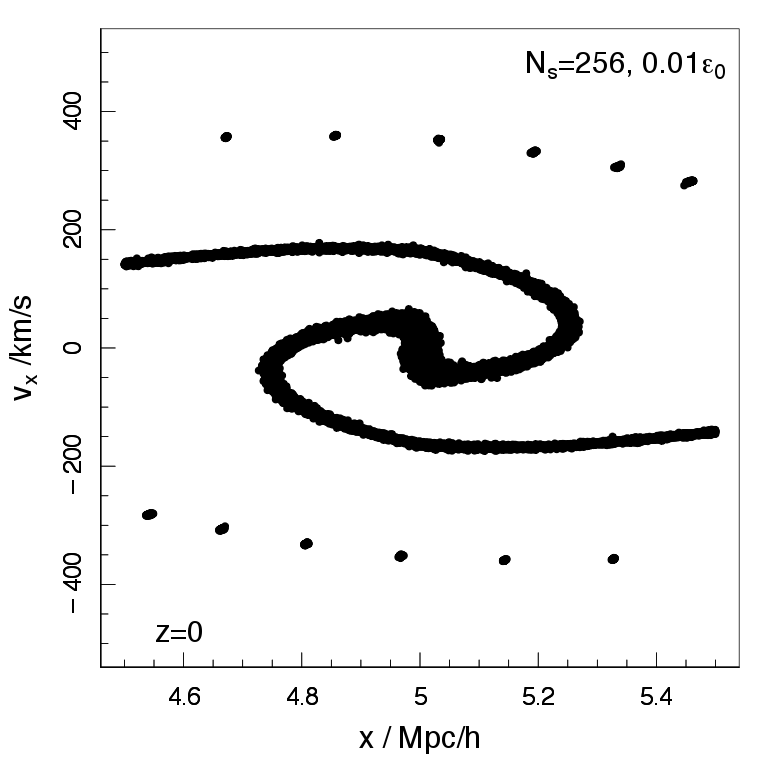}}
  
  \caption{{\bf Plane-Symmetric Collapse: Phase Space Structure.} Here we
    show the phase space structure in the $256^3$ run, at approximately
    shell crossing ($z$=4; upper panels) and at late times ($z$=0; lower
    panels). Here $v_x$ is the peculiar velocity along the $x$-direction and
    $x$ is the comoving position. The gravitational softening $\epsilon$
    decreases by factors of 10 from $\epsilon_0$ (left-most panel)
    to 0.01 $\epsilon_0$ (right-most panel).}
  \label{fig:pancake_phase}
\end{figure*}

\paragraph*{Phase Space Structure:} In Figure~\ref{fig:pancake_phase} we
show slices through phase space of the one-dimensional wave (peculiar velocity
$v_x$ versus comoving position $x$) at approximately shell crossing ($z\simeq4$,
upper panels) and at late times ($z\simeq 0$, lower panels). Here we have
varied systematically the gravitational comoving softening, adopting values of
(from left to right) $\epsilon/\epsilon_0$ = (1, 0.1, 0.01) in the $256^3$ run.
What is striking in this Figure is the impact that
$\epsilon$ has on the phase space structure of the wave. At early times, prior
to shell crossing, the difference in the evolution of the waves in the
different runs is negligible. It is at shell crossing that differences in
evolution start to become apparent -- the smaller the value of
$\epsilon$, the more pronounced the deviation from the predicted
evolution. Rather than evolving as a thin sheet that preserves its structure
as it winds up (as in the cases when $\epsilon=\epsilon_0$), the sheet
thickens at shell crossing, arising from velocity perturbations that grow
with decreasing $\epsilon$ as $\epsilon^{-1}$ (cf. Eq~\ref{eq:deltav})
and which fragment into clumpy structures in phase space at later times
(cf. lower-middle and lower-right panels).

\begin{figure*}
  \centerline{\includegraphics[width=5.3cm]{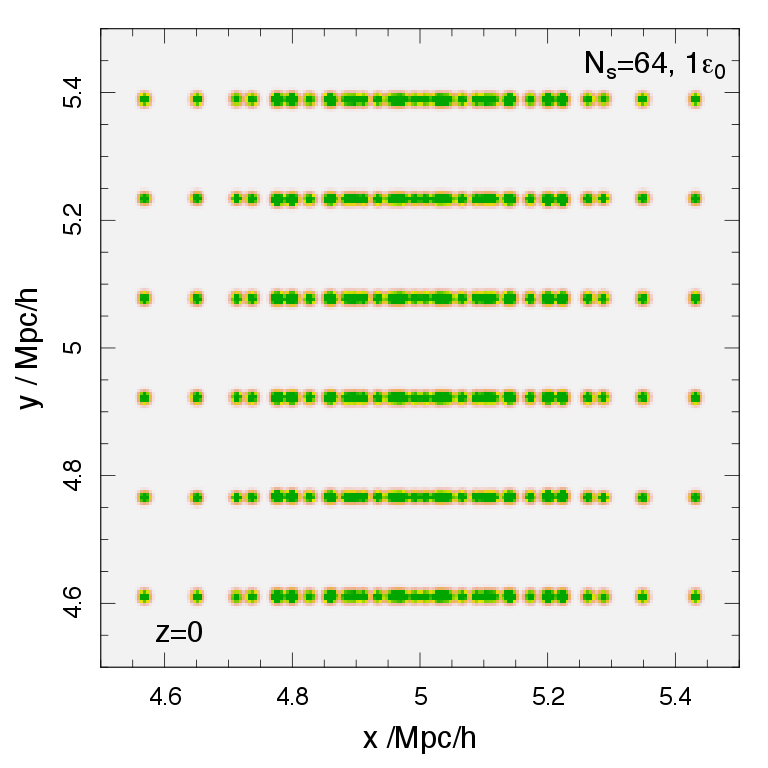}
      \includegraphics[width=5.3cm]{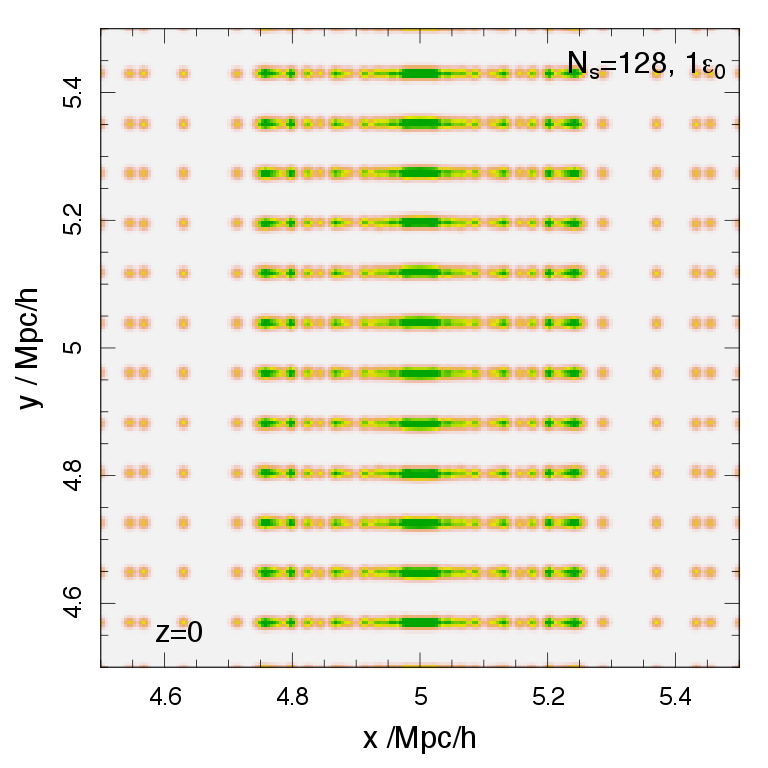}
      \includegraphics[width=5.3cm]{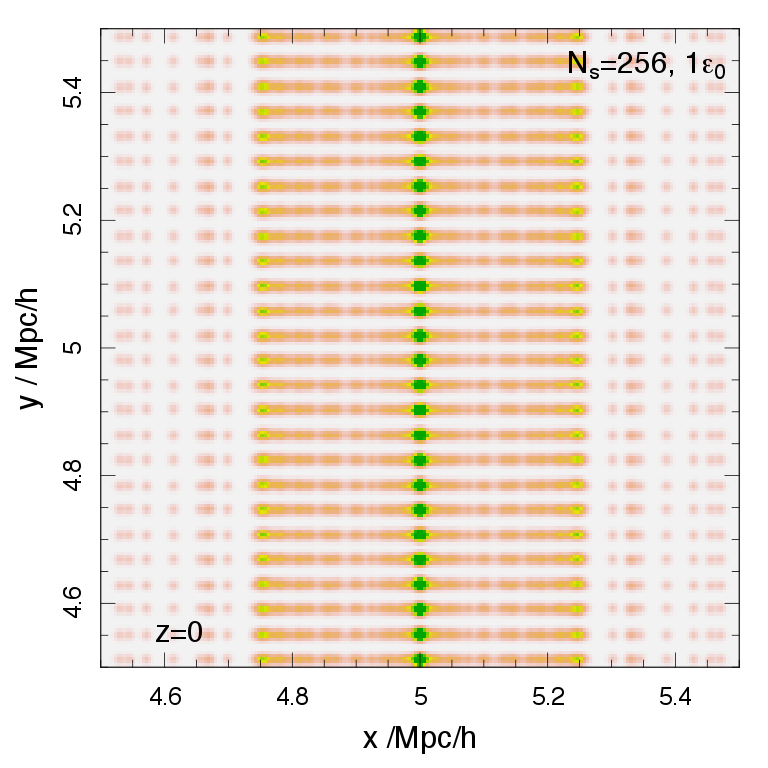}}
  \centerline{\includegraphics[width=5.3cm]{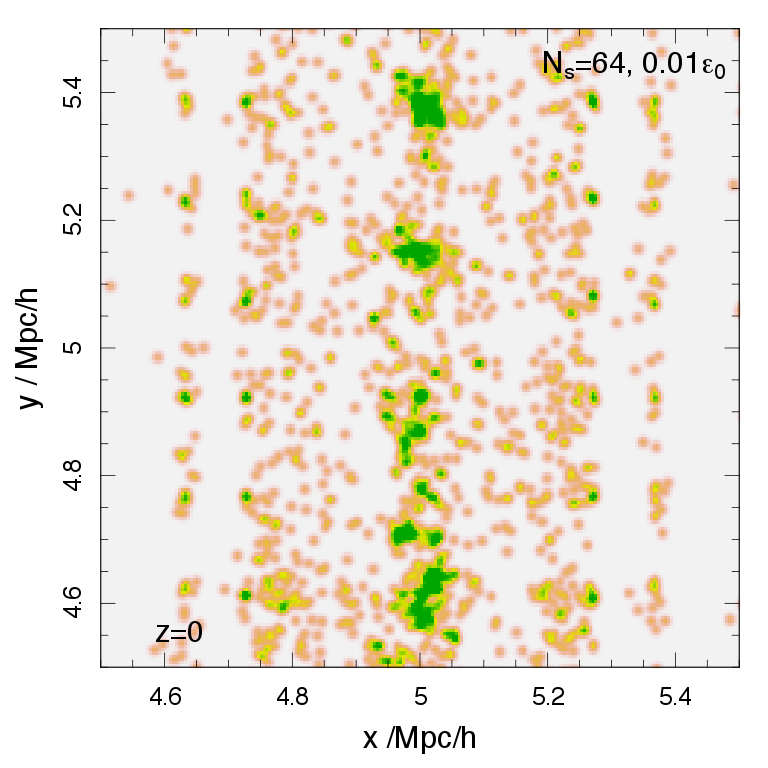}
    \includegraphics[width=5.3cm]{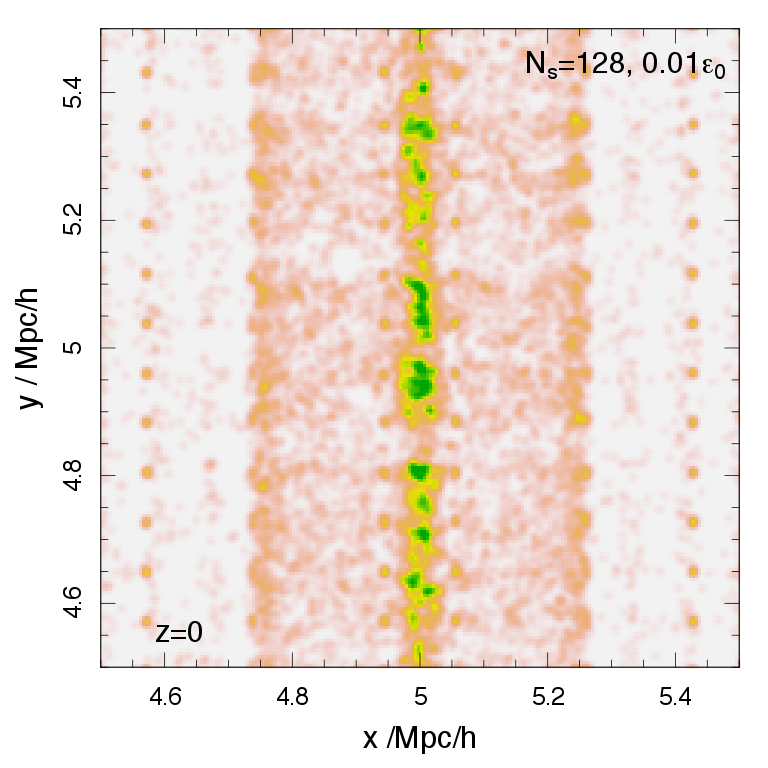}
    \includegraphics[width=5.3cm]{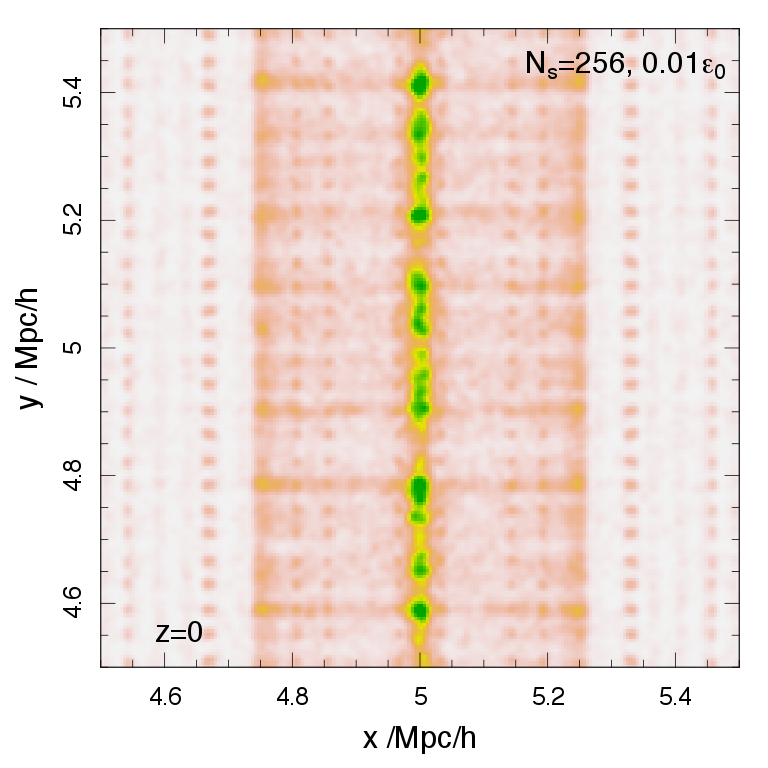}}

  %  \centerline{\includegraphics[width=5.3cm]{./xy051_256_eps1.pdf}
  %    \includegraphics[width=5.3cm]{./xy051_256_eps01.pdf}
  %    \includegraphics[width=5.3cm]{./xy051_256_eps001.pdf}}
  \caption{{\bf Plane-Symmetric Collapse: Spatial Structure.} Here we show 
    the projected spatial distribution (in the $x$-$y$ plane) at late times
    ($z$=0) in the $64^3$, $128^3$ and $256^3$ runs (left to right) with
    gravitational softening lengths $\epsilon$ = 0.01 $\epsilon_0$, where
    $\epsilon_0$ = $\bar{d}$.}
  \label{fig:pancake_space}
\end{figure*}

\paragraph*{Spatial Structure:} In Figure~\ref{fig:pancake_space} we show the
resulting projected spatial distribution ($x$ versus $y$) of particles at
$z$=0 within the region where the density enhancement is greatest in the
$64^3$, $128^3$ and $256^3$ runs (left to right columns) for adopted
softening of $\epsilon/\epsilon_0$=1 (top row) and 0.01 (bottom row).
The smaller value of $\epsilon$ is a little more conservative than is usually
adopted in large uniform resolution $N$-body simulations (a factor of $3-5$
smaller), but it highlights the point that we wish to make. We expect particles
to oscillate collisionlessly about the midplane ($x$=5$h^{-1} \rm Mpc$); this
is what we observe in the regularity of the $\epsilon$=$\epsilon_0$ case, with
particle density varying smoothly along their trajectory. This regularity
breaks down as $\epsilon$ is reduced and distinct clumps have formed in the
midplane, which are spaced at roughly the mean inter-particle separation of
the simulation. This is what we would expect based on the behaviour evident
in Figure~\ref{fig:pancake_phase} -- if $\epsilon$ is too small, $N$-body
particles are subject to large velocity perturbations in what would otherwise
be the smooth gravitational field in which they move because of close
encounters with other $N$-body particles. These perturbations act to
scatter particle momenta, diffusing the initial planes away
and producing the clumps evident in the midplane in the lower panels columns
of Figure~\ref{fig:pancake_space}.

\begin{figure*}
  \centerline{
    \includegraphics[width=5.3cm]{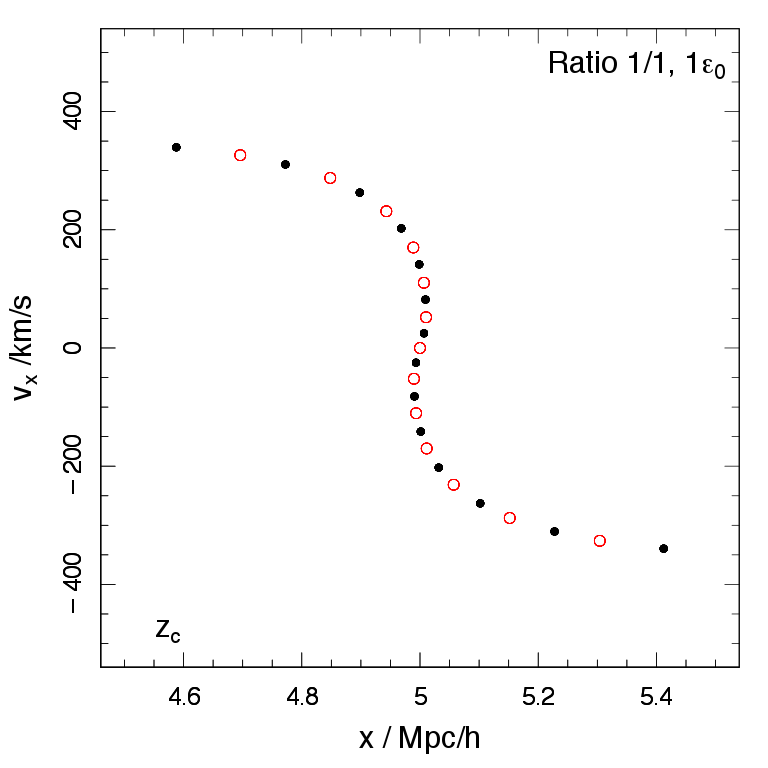}
    \includegraphics[width=5.3cm]{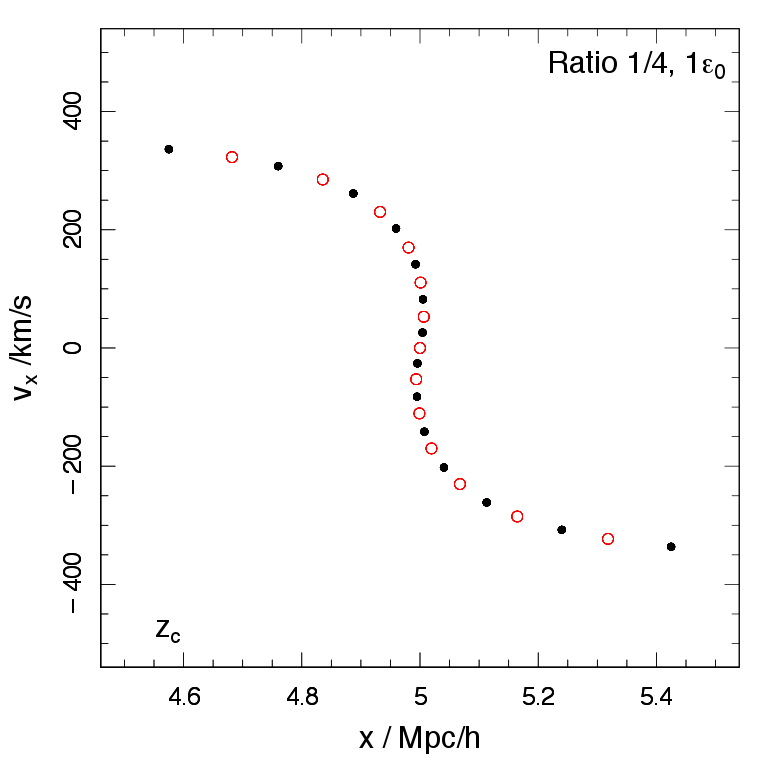}
    \includegraphics[width=5.3cm]{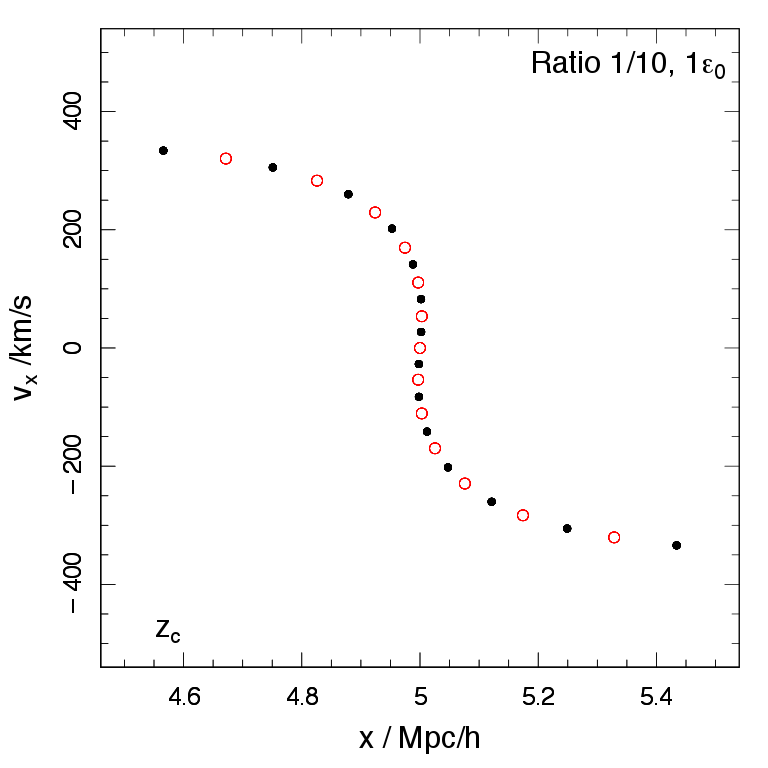}}
  \centerline{
    \includegraphics[width=5.3cm]{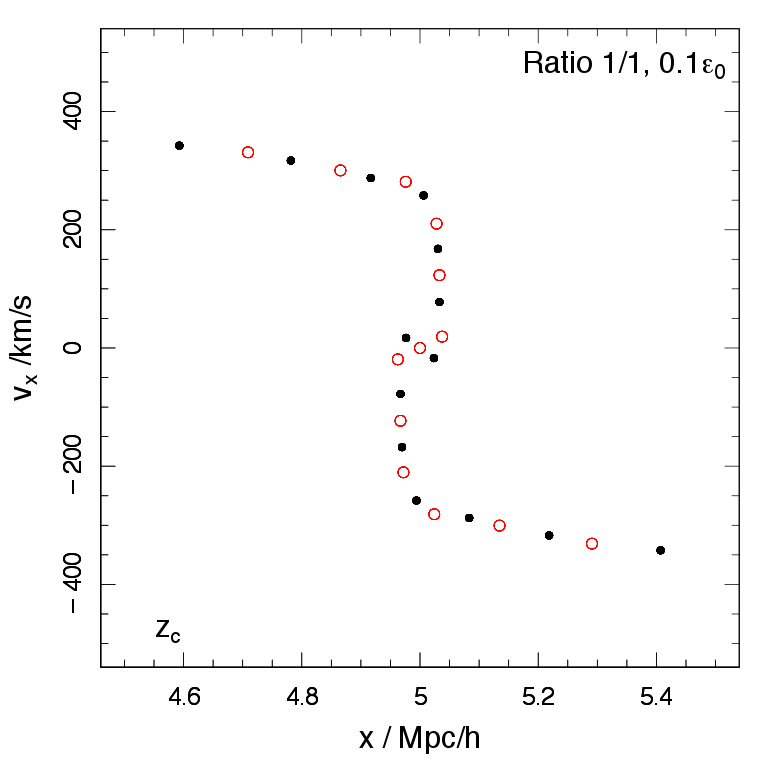}
    \includegraphics[width=5.3cm]{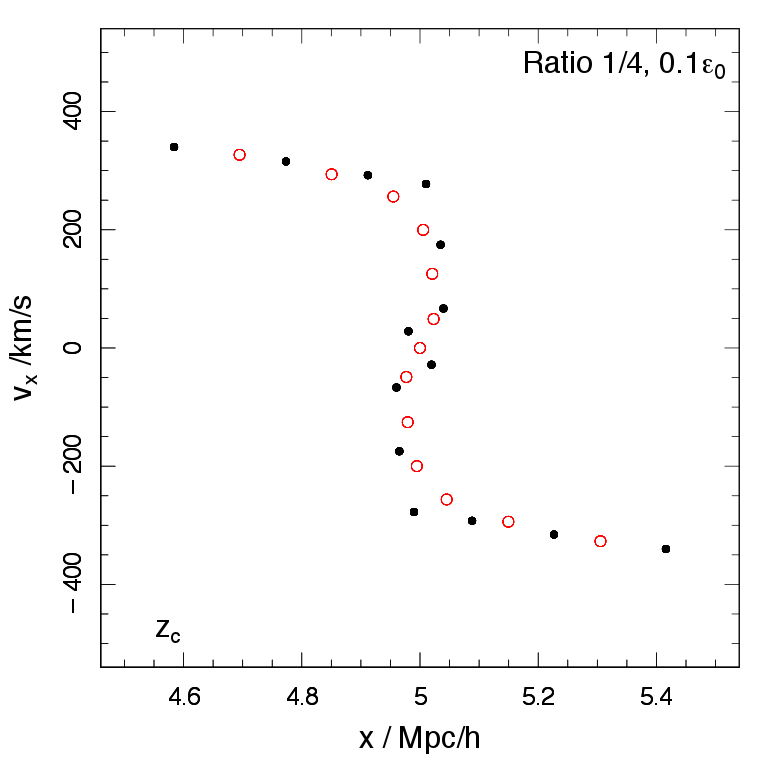}
    \includegraphics[width=5.3cm]{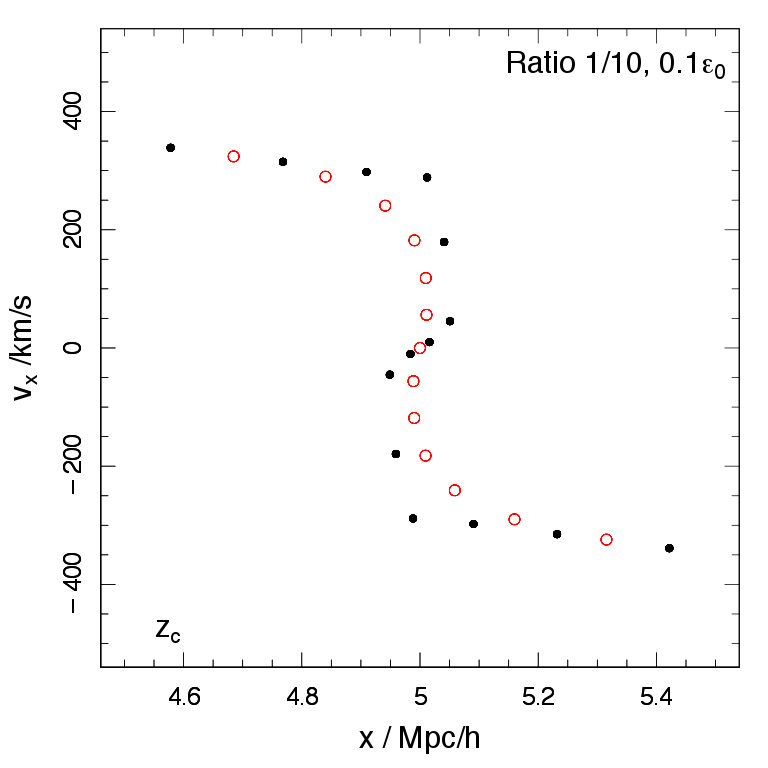}}
  \centerline{
    \includegraphics[width=5.3cm]{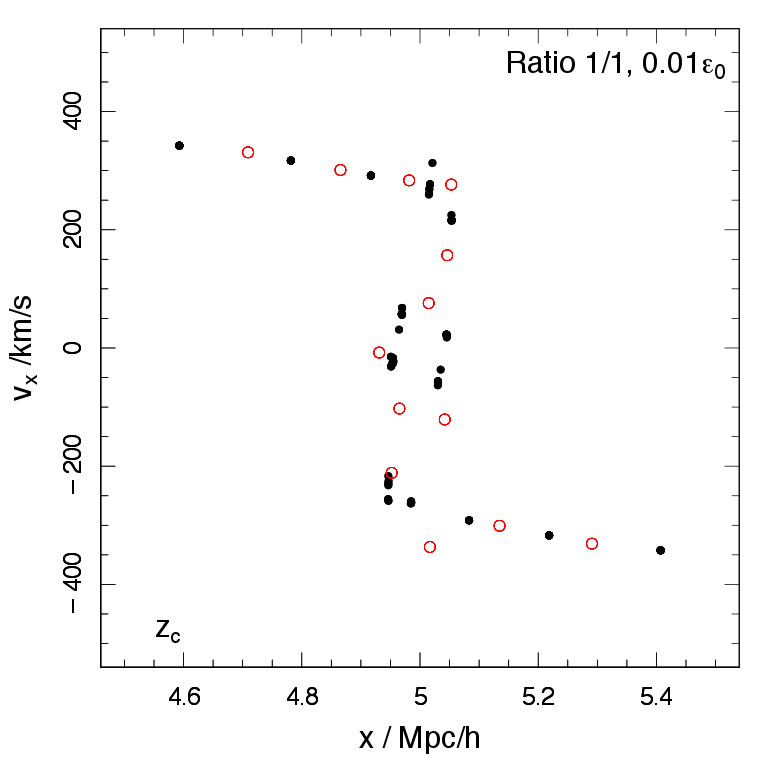}
    \includegraphics[width=5.3cm]{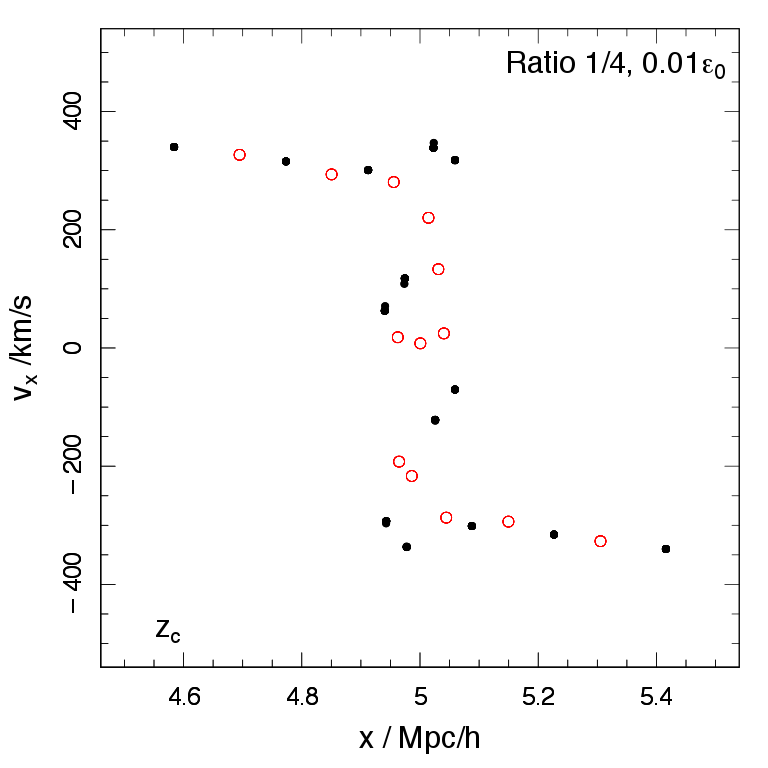}
    \includegraphics[width=5.3cm]{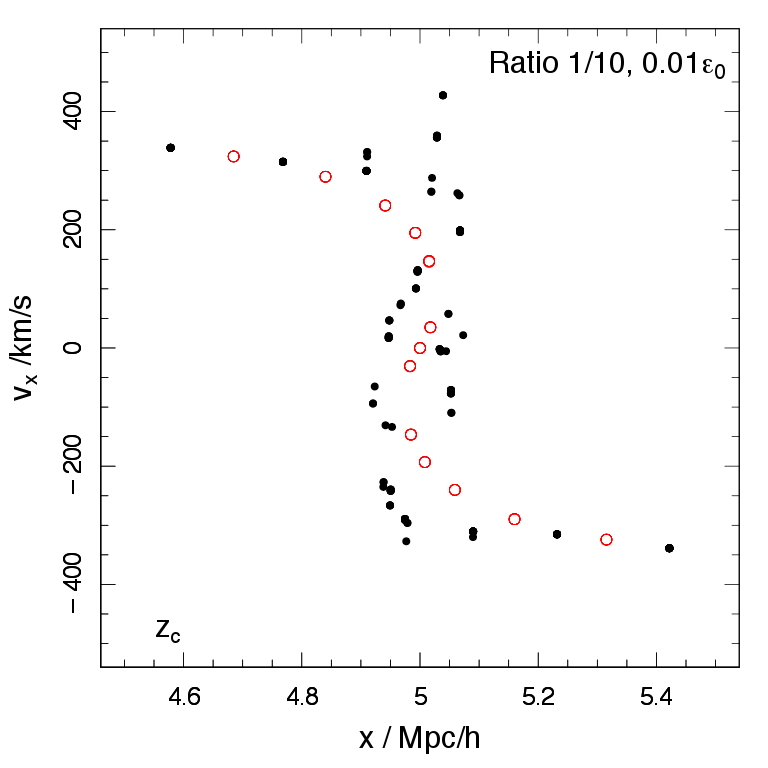}}  
  \caption{{\bf Two-Component Plane-Symmetric Collapse: Phase Space Structure} 
    In the upper, middle, and lower rows, we show the phase space structure
    at $z_c \simeq 4$ in the $\epsilon/\epsilon_0$=(0.01,0.1,1.) runs, where
    the ratio of particle masses varies from 1/1 (left column) to 1/4
    (middle column) to 1/10 (right column). Filled (open) circles correspond
    to the less (more) massive component.}
  \label{fig:two_species_pancakes}
\end{figure*}

\begin{figure*}
  \centerline{
    \includegraphics[width=5.3cm]{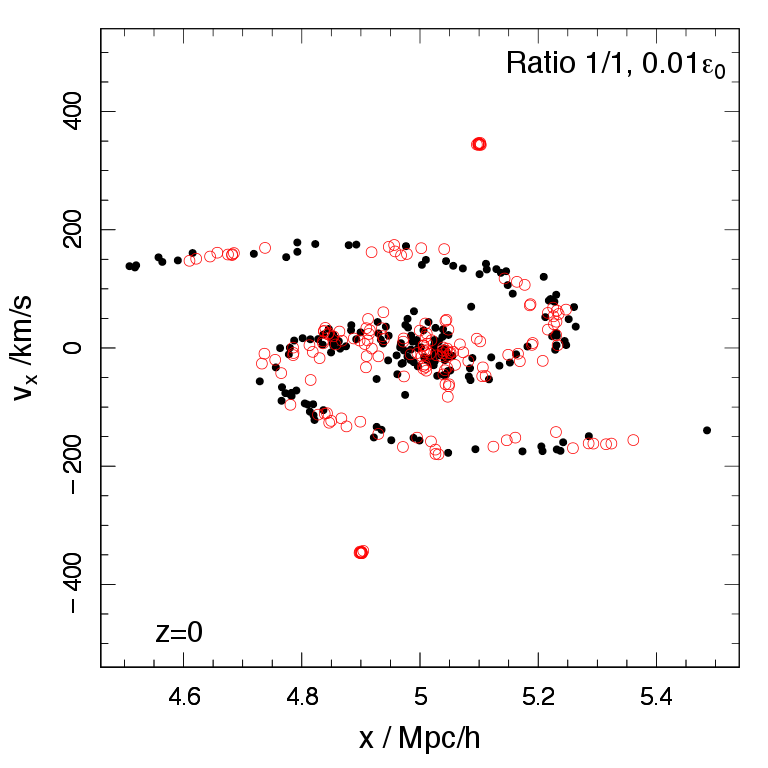}
    \includegraphics[width=5.3cm]{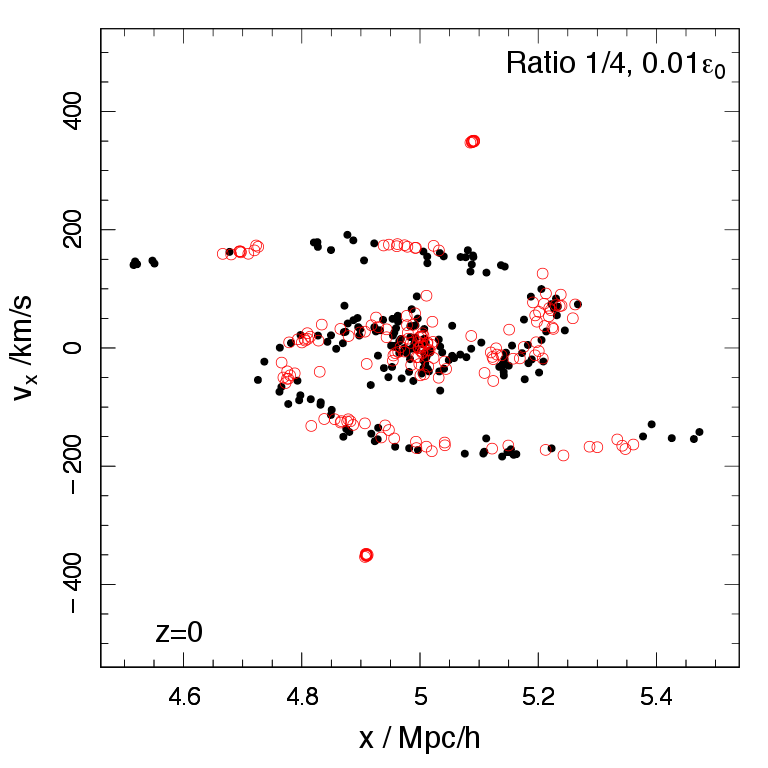}
    \includegraphics[width=5.3cm]{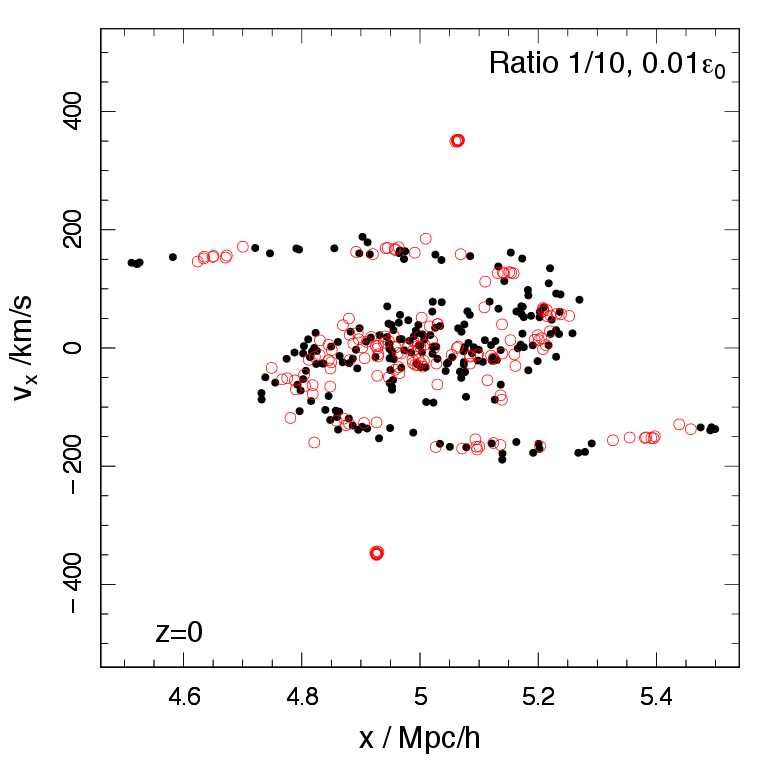}}
  \caption{{\bf Two-Component Plane-Symmetric Collapse: Phase Space Structure} 
    Here we show the phase space structure at $z$=0 in the $\epsilon/\epsilon_0$=0.01
    run, where the ratio of particle masses varies from 1/1 to 1/4 to 1/10 (left to right).
    Filled (open) circles correspond to the less (more) massive component.}
  \label{fig:two_species_pancakes_z0}
\end{figure*}

\paragraph*{Two-Component Pancakes:} To verify behaviour we see in the
cosmological simulations present in \S~\ref{ssec:cosmosims}, we have also
run two collisionless components versions of these experiments, looking at
mass ratios of $(1/1,1/\sqrt{2},1/4,1/10)$ and softenings of
$\epsilon/\epsilon_0$=(1, 0.1, 0.01). The results are consistent with those
already presented -- we see systematic deviations from the expected evolution
for softenings $\epsilon\ll\epsilon_0$ and these are seeded at shell crossing.
As highlighted by Figure~\ref{fig:two_species_pancakes}, which shows the phase
space structure in these two-component pancakes at approximate shell crossing
($z_c \simeq 4$), that the larger the mass ratio, the larger the initial
momentum perturbation experienced by the less massive collisionless component
(filled circles), and as the system evolves post-shell crossing, the more
the two components mix in the area of overlap (see
Figure~\ref{fig:two_species_pancakes_z0}).

We note at this point that these trends are consistent with behaviour
reported by \citet{angulo.2013a}, who found evidence for spurious coupling
between baryon and dark matter components in their cosmological simulations
-- where the two components have a mass ratio of $\sim\!1/6$. They argued
that this arises from a loss of collisionality that arises from
discretization of the density field and the use of high force resolution
(i.e. $\epsilon\ll\epsilon_0$), and can be overcome by having ``softer''
gravitational forces on scales below the mean inter-particle separation,
to suppress spurious coupling between baryons and dark matter. We return to
this point below.

\subsection{Cosmological Simulations}
\label{ssec:cosmosims}

To assess how discreteness-driven relaxation of the form just described
affects cosmological $N$-body simulations, we follow
\citet[][hereafter BK02]{binney.knebe.2002} and use two sets of collisionless
particles to represent the dark matter distribution. BK02 were
interested in identifying the influence of two-body relaxation on the
internal structure of dark matter haloes by looking for evidence of mass
segregation, with the more massive component's particles preferentially
occupying the inner parts of haloes. Here we are interested in how more
and less massive (hereafter heavier and lighter) $N$-body particles cluster,
with the expectation that differences should be negligible if discreteness
effects are unimportant. In what follows, we show results for the runs in
which the mass ratio is $1/4$, but have also checked for consistency in
trends present in the runs with a mass ratio of $1/\sqrt{2}$. We also
distinguish between FOF groups, overdensities recovered by the algorithm in
the dark matter density field that can be either physical or spurious, rather
than haloes, the subset of FOF groups that correspond to physical overdensities.

\begin{figure*}
  \centerline{\includegraphics[width=5.3cm]{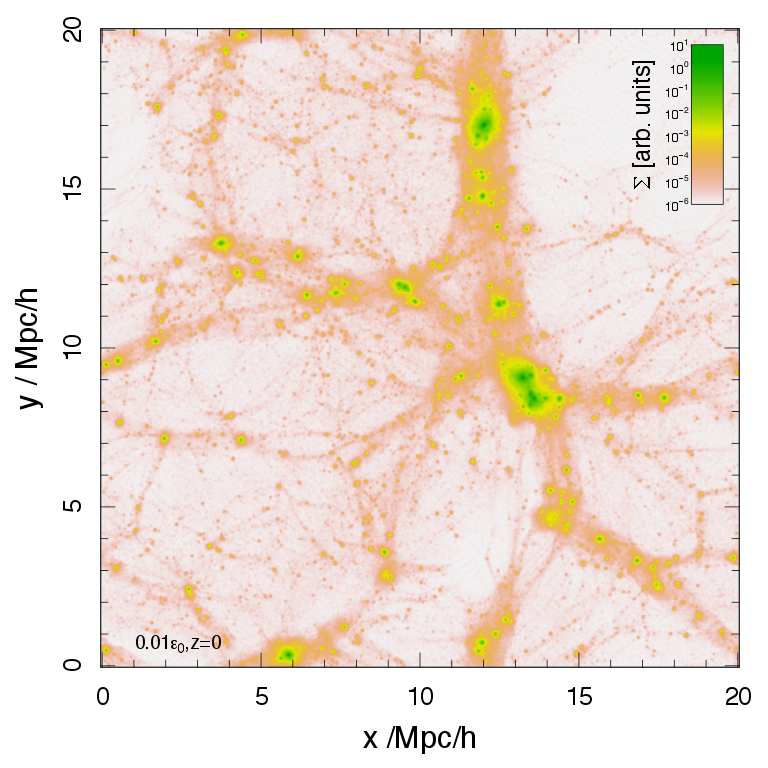}
    \includegraphics[width=5.3cm]{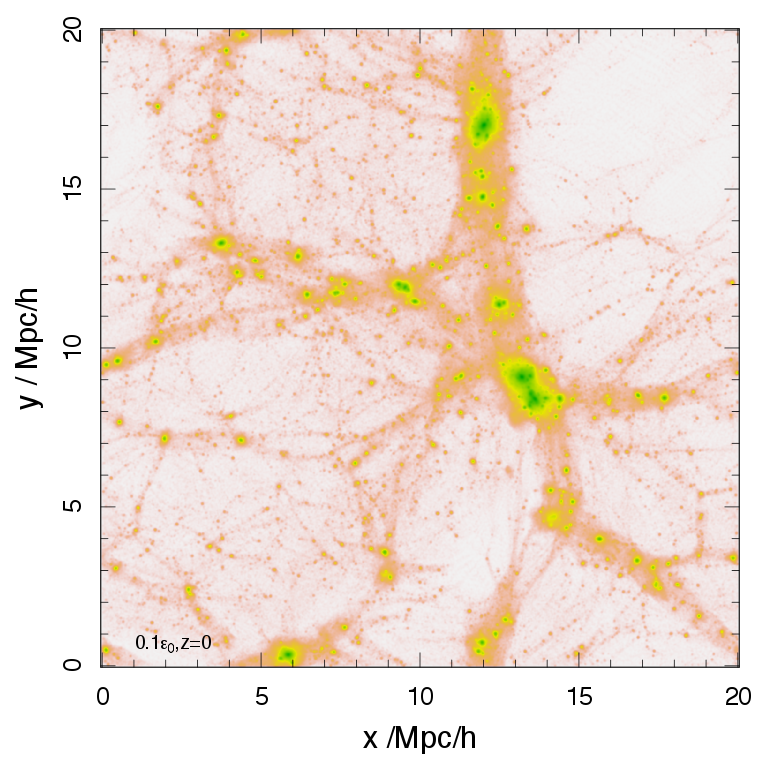}
    \includegraphics[width=5.3cm]{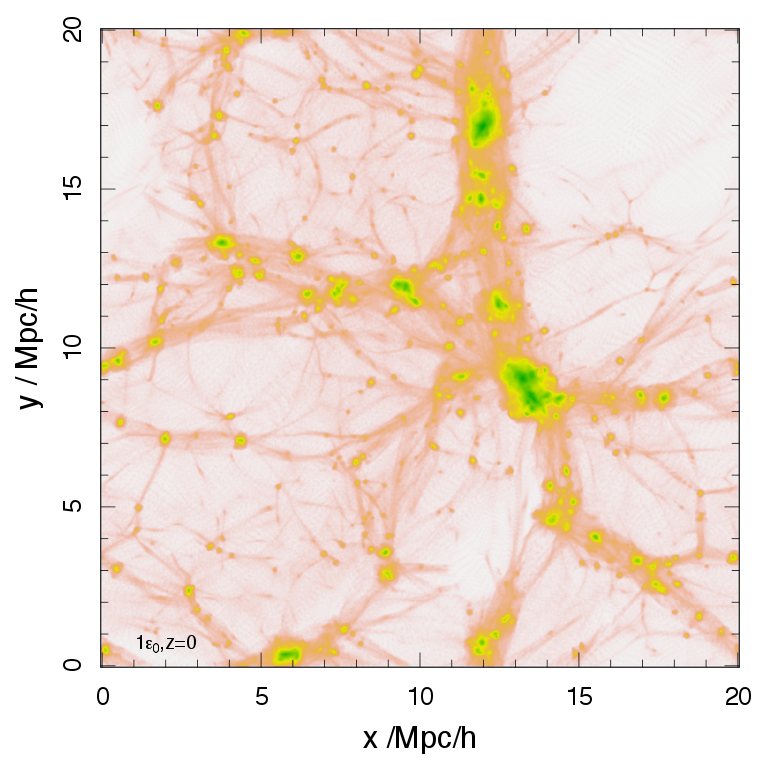}}
  \centerline{
    \includegraphics[width=5.3cm]{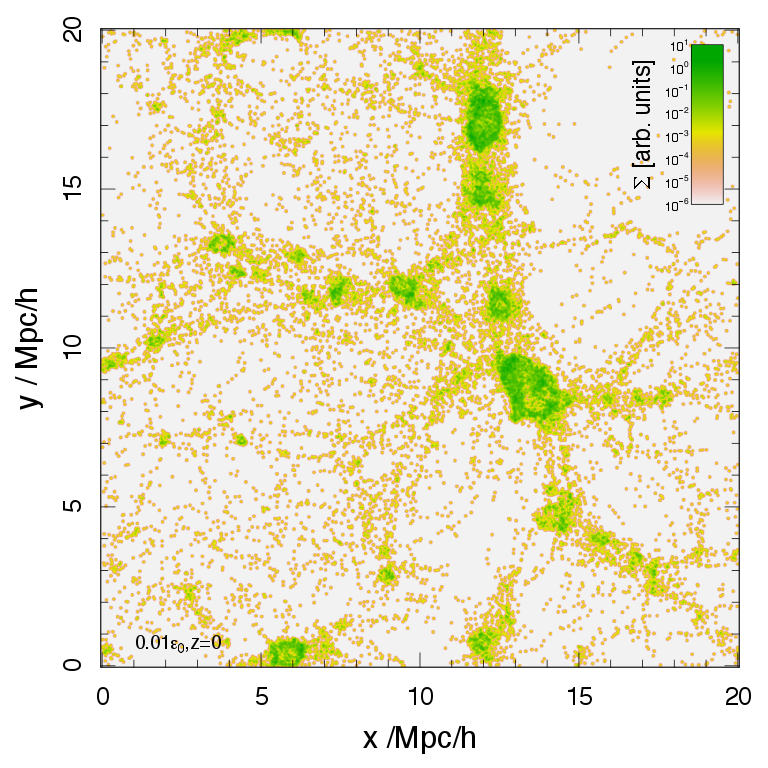}
    \includegraphics[width=5.3cm]{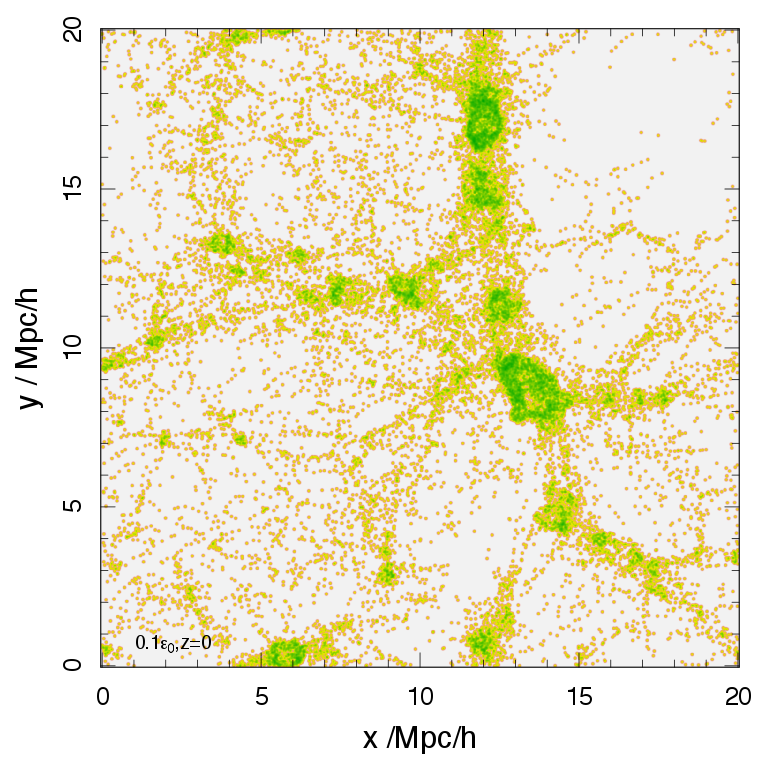}
    \includegraphics[width=5.3cm]{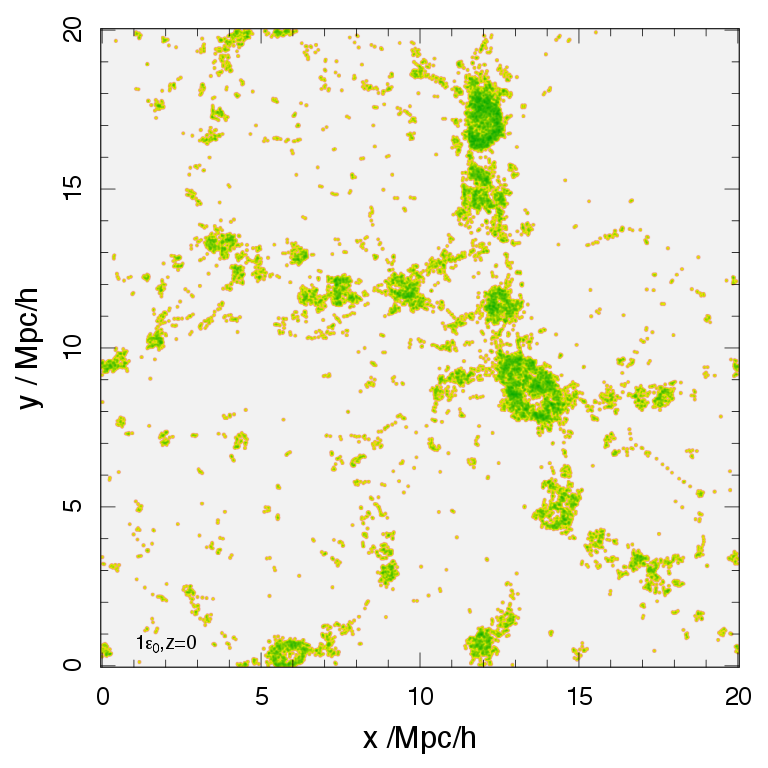}}

  \caption{{\bf Spatial Distribution of FOF Groups: CDM Case.} 
    Upper Panels: The projected dark matter density distribution in the 
    $\epsilon/\epsilon_0$=(0.01,0.1,1.) runs. The colour bar indicates
    the projected density scale in arbitrary units; the same scale is used
    in each panel.
    Lower Panels: Spatial
    distribution of FOF group centres, where centre corresponds to
    centre of density $\vec{r}_{\rm cen}$; see text for further details.}
  \label{fig:cdm_haloes}
\end{figure*}

\begin{figure*}
  \centerline{\includegraphics[width=5.3cm]{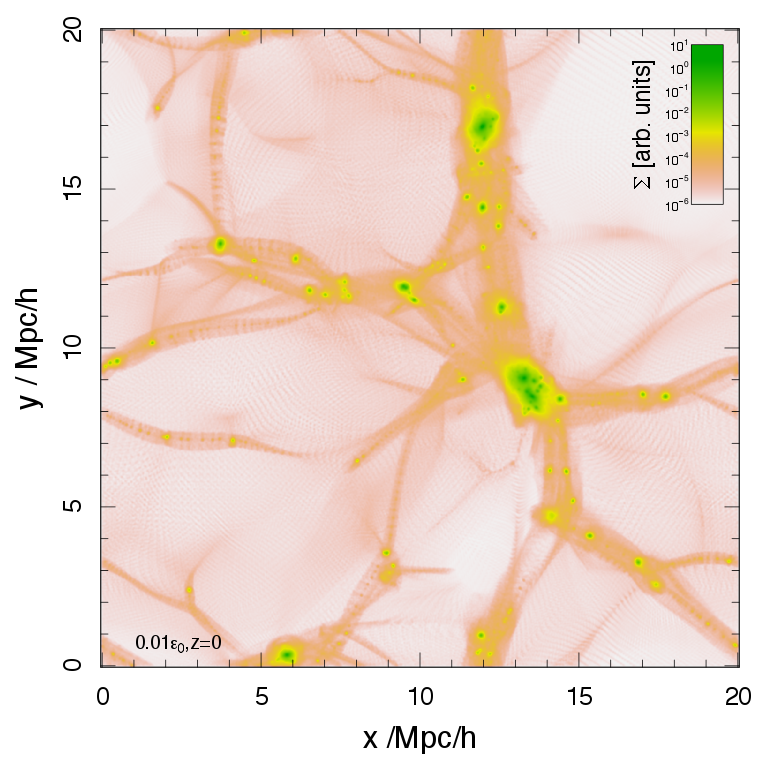}
    \includegraphics[width=5.3cm]{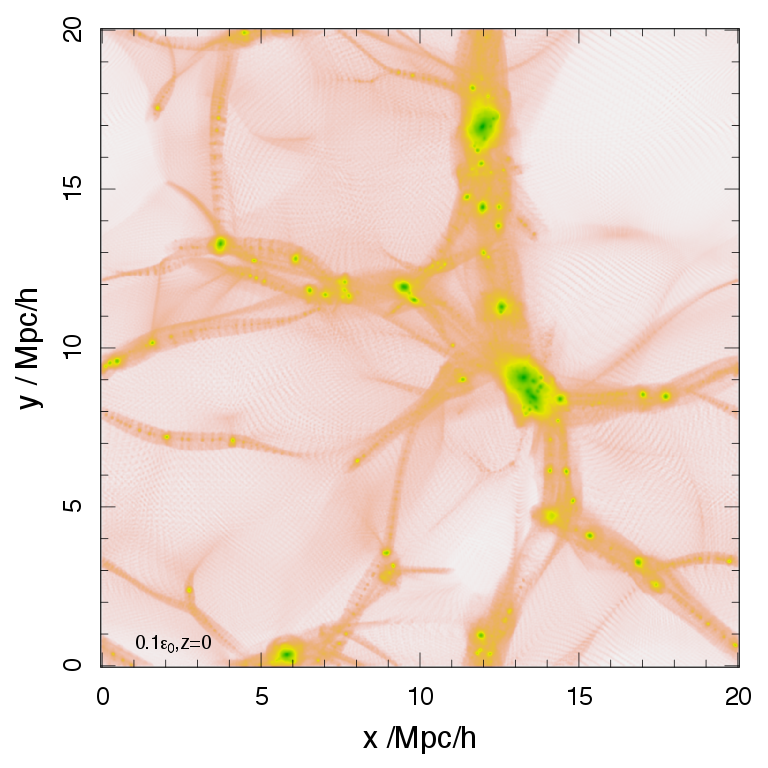}
    \includegraphics[width=5.3cm]{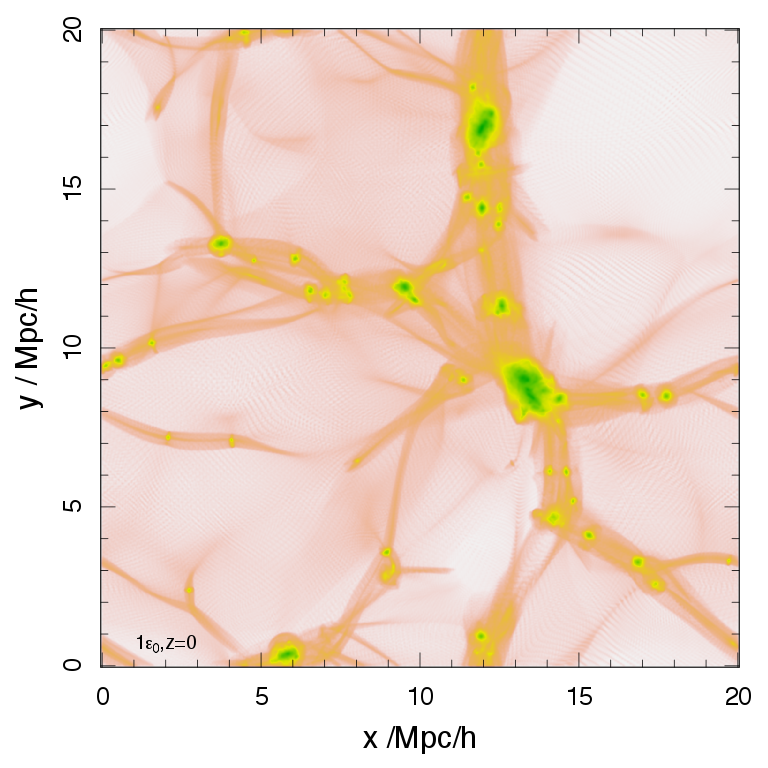}}

  \centerline{
    \includegraphics[width=5.3cm]{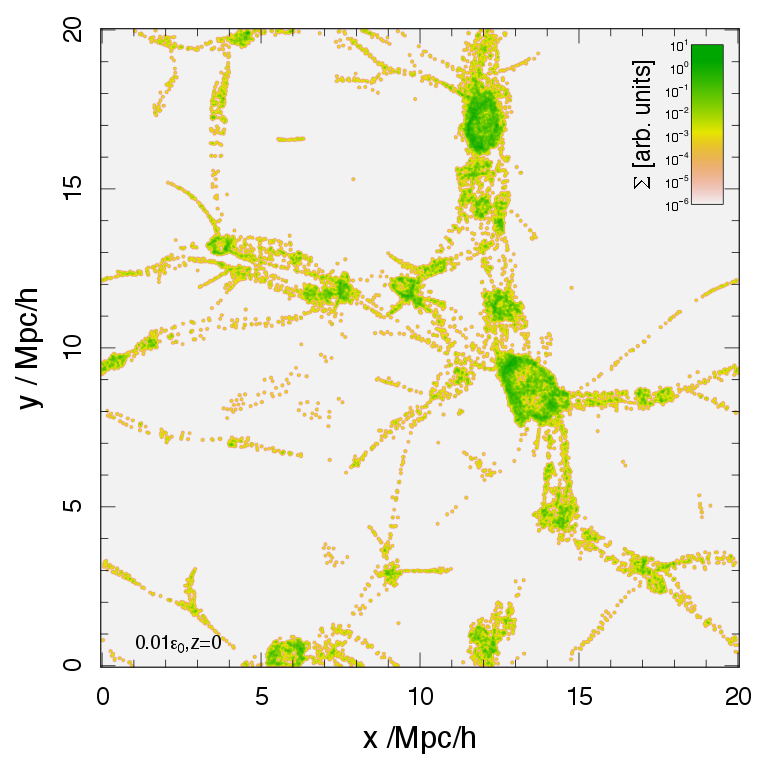}
    \includegraphics[width=5.3cm]{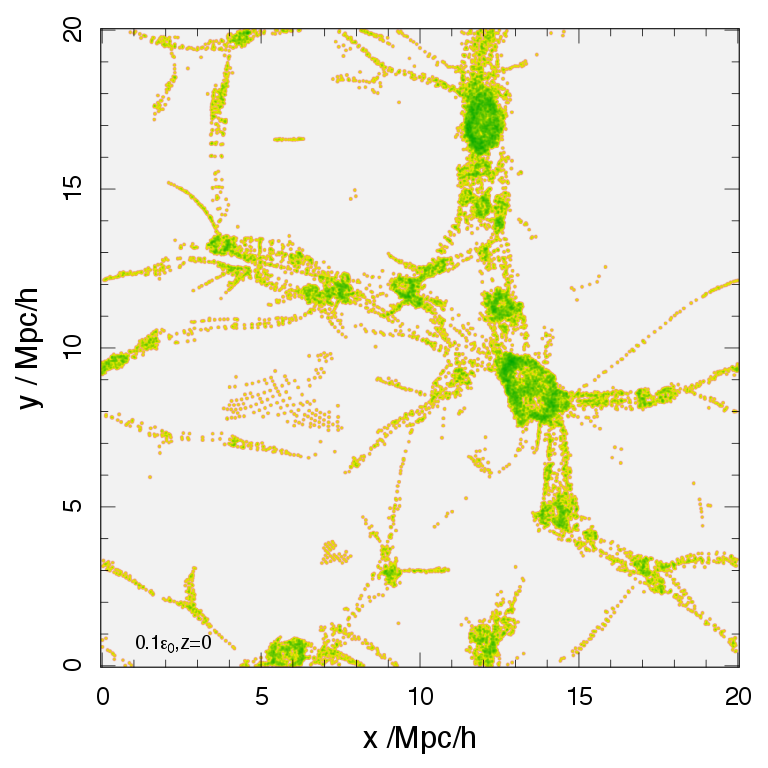}
    \includegraphics[width=5.3cm]{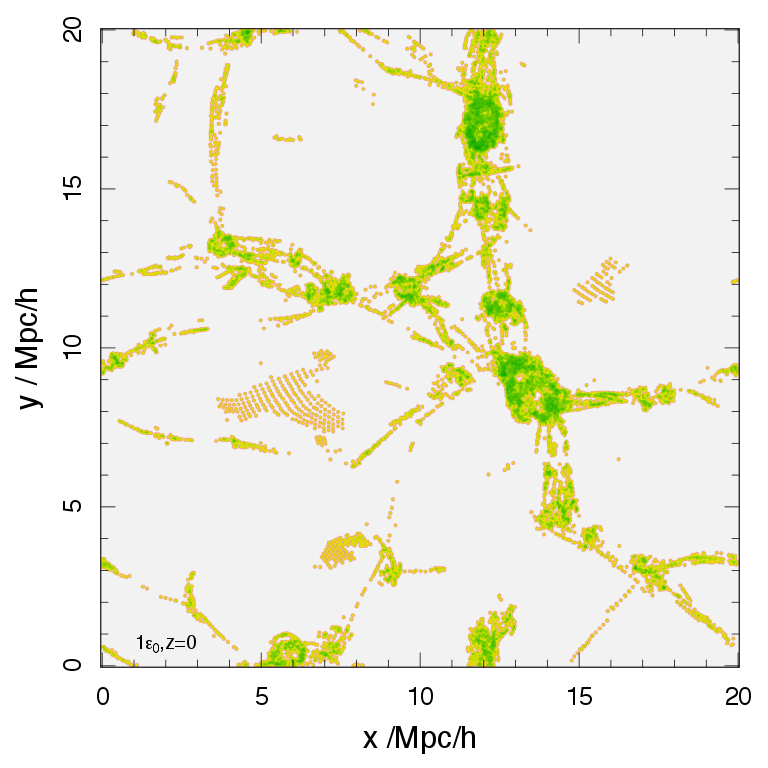}}
  \caption{{\bf Spatial Distribution of FOF Groups: WDM Case.} 
    As in Figure~\ref{fig:cdm_haloes}, we show the projected dark matter 
    density distribution in the $\epsilon/\epsilon_0$=(0.01,0.1,1.) runs
    (upper panels) and the spatial distribution of FOF group
    centres (lower panels).}
  \label{fig:wdm_haloes}
\end{figure*}

\paragraph*{Spatial Distribution:} In Figures~\ref{fig:cdm_haloes}
and~\ref{fig:wdm_haloes} we show the $z$=0 projected dark matter density
distribution (upper panels) and the locations of the centres of FOF groups
with $N_{\rm FOF}\geq 30$ particles (lower panels) within a 2 $h^{-1}$ Mpc slice
in the CDM and WDM runs (lower panels), evolved with (from left to right)
$\epsilon/\epsilon_0$=(0.01,0.1,1);
the colour bar indicates logarithm of projected $N$-body particle/FOF group
number
density. These Figures show that the properties of the large-scale structure
-- its topology, the locations of massive groups, and the density contrast
within the dominant filaments -- are consistent between CDM and WDM runs and
between runs with different softenings, and the deficit in the abundance of
small-scale structure in the WDM runs with respect to the corresponding CDM
run is also readily apparent.

However, they also reveal some interesting differences in the influence of
$\epsilon$ on the structure of filaments and the distribution of FOF groups
between the CDM and WDM runs. In the WDM runs, varying $\epsilon/\epsilon_0$
between 0.01 to 1 has a relatively small impact on either large-scale
filamentary structure or the spatial distribution of groups -- filaments
delineated by groups are readily identifiable in all of the runs. The
projected dark matter density within filaments in the $\epsilon/\epsilon_0$=1
run is smoother than in the $\epsilon/\epsilon_0<1$ runs, although the
projected FOF group distribution reveals the presence of low-mass groups.
Small differences are evident in the low-density regions in the projected
distribution of FOF groups, where the artifact of the initial particle grid
has been picked up by the FOF algorithm, most noticeably in the
$\epsilon/\epsilon_0$=1 case centred on
$(x,y)\simeq(6,8)h^{-1} \rm Mpc$. In contrast, the impact of $\epsilon$
is more marked in the CDM runs than in the WDM runs, with
the key differences being between the $\epsilon/\epsilon_0<$1 and
$\epsilon/\epsilon_0$=1 cases. Contrasting these two cases and
focusing on the projected dark matter density maps, we see that
minor filaments are smooth and continuous when $\epsilon/\epsilon_0$=1
whereas they are fragmented and clumpy when $\epsilon/\epsilon_0<$1, while
low density regions contain fewer structures when $\epsilon/\epsilon_0$=1
compared to the $\epsilon/\epsilon_0<$1 case. The projected FOF group
distribution reveals a fog of low-mass groups filling the simulation volume when
$\epsilon/\epsilon_0<$1, whereas this fog is suppressed when
$\epsilon/\epsilon_0$=1; the density of lower-mass groups in the environs of
the most massive groups within the simulation volume is similar for all values
of $\epsilon$.

\begin{figure}
  \centerline{\includegraphics[width=0.99\columnwidth]{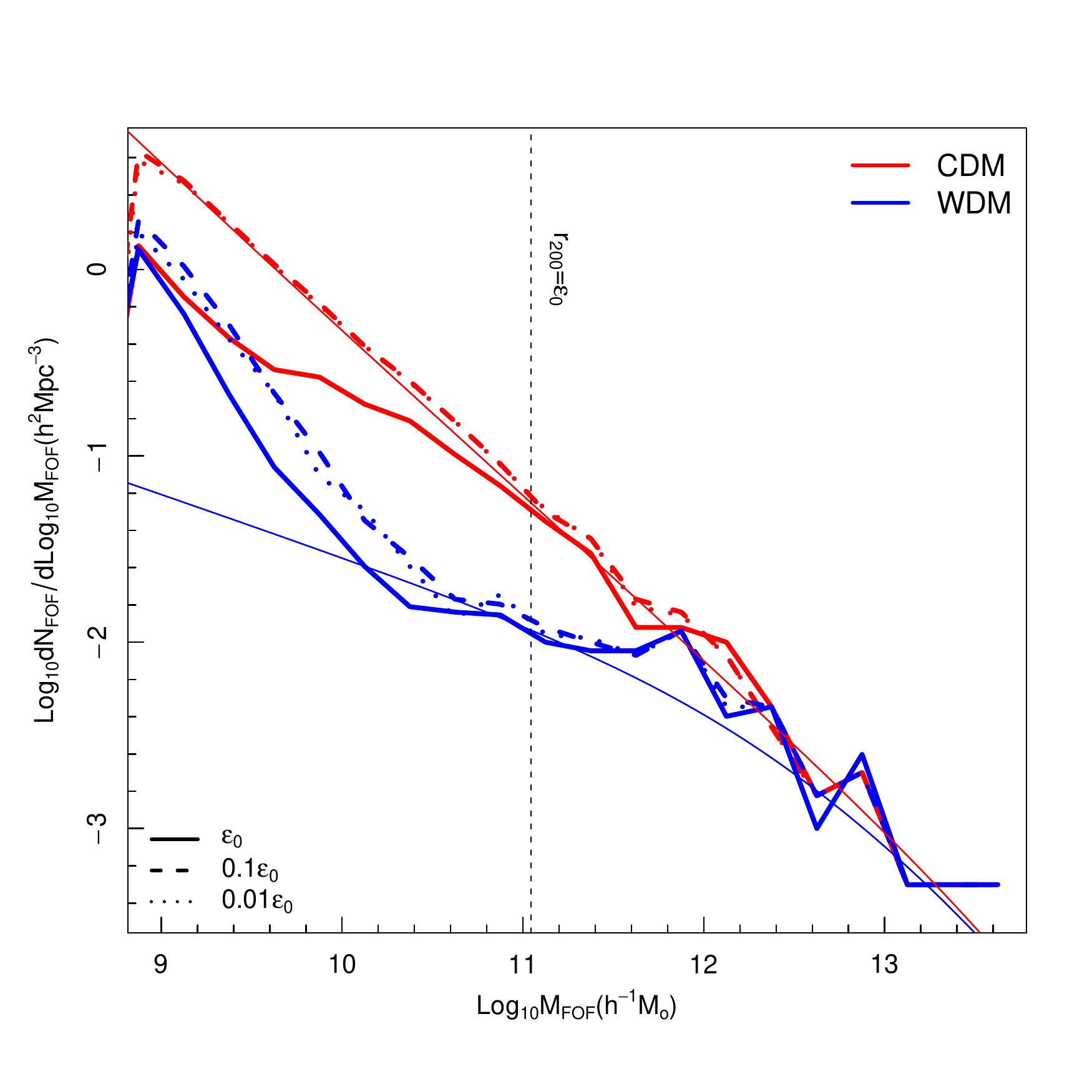}}
  \caption{{\bf FOF Group Mass Functions.} Here we show the differential
    mass functions of FOF groups (containing both components of particle) in
    the $\epsilon/\epsilon_0$=(1,0.1,0.01) runs (heavy solid, dashed, dotted);
    light solid curves indicate the appropriate CDM and WDM halo mass
    functions, computed using {\small{HMFcalc}} \citep{murray.etal.2013},
    while the vertical dashed line indicates the (approximate) minimum halo
    mass we might expect in the $\epsilon/\epsilon_0$=1 run,
    where the halo virial radius $r_{200}$ is set to $\epsilon_0$.}
  \label{fig:mass_func_comp}
\end{figure}

\begin{figure*}
  \centerline{\includegraphics[width=0.99\columnwidth]{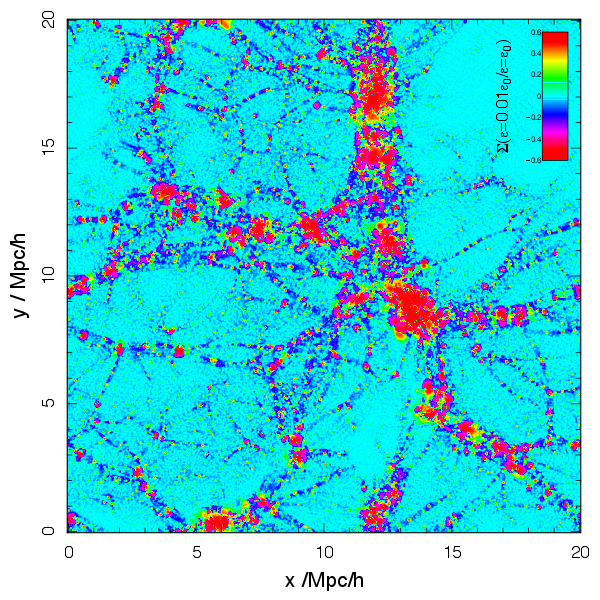}
    \includegraphics[width=0.99\columnwidth]{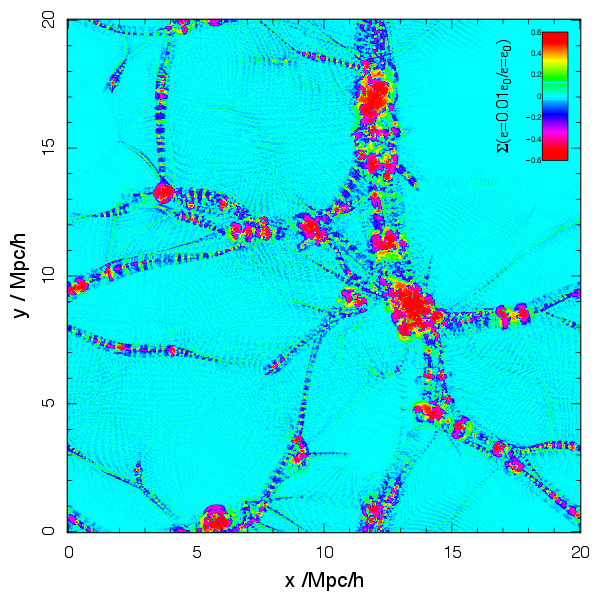}}
  \caption{{\bf Difference Maps: Delineating by Softening.} Here we show
    the difference in projected density distribution between the
    $\epsilon$=0.01$\epsilon_0$ and $\epsilon$=$\epsilon_0$ CDM and WDM
    runs (left and right panels respectively). We use the {\tt atan}
      mapping in the {\small R} {\tt magicaxis} package.}
  \label{fig:differences}
\end{figure*}

We quantify these visual impressions in Figures~\ref{fig:mass_func_comp}
and~\ref{fig:differences}, where we make explicit how varying $\epsilon$
in softening affects structure in the CDM and WDM runs.
Figure~\ref{fig:mass_func_comp} shows the differential mass functions measured
in the CDM (red curves) and WDM (blue curves) runs for softenings of
$\epsilon/\epsilon_0$=1 (solid), 0.1 (dashed), and 0.01 (dotted), where
masses correspond to $M_{\rm FOF}$ rather than $M_{200}$. For
comparison, \citet{sheth.tormen.1999} mass functions estimated using
the {\small{HMFcalc}} tool of \citet{murray.etal.2013} for the appropriate
cosmological dark matter model are shown as light curves. The vertical dashed
line indicates the halo mass corresponding to $r_{200}=\epsilon_0$, i.e.
$M(\epsilon_0)=4\pi/3 \times 200 \times \rho_{\rm crit} \times \epsilon_0^3$,
which we might consider to be (approximately) the minimum halo mass in the
$\epsilon/\epsilon_0$=1 runs.

Choosing $\epsilon/\epsilon_0$=1 suppresses the mass function for
$M_{\rm FOF}\lesssim 10^{11}h^{-1} \rm M_{\odot}$ in both CDM and WDM runs,
but there is no sharp decline with decreasing mass in either case, as we
might expect; indeed, the upturn in the mass function associated with WDM
mass functions persists, albeit shifted to smaller masses by $\sim 0.2$ dex.
Choosing $\epsilon/\epsilon_0$=0.1 or 0.01 makes little difference to either
the shape or amplitude of the mass function. This persistence of the
upturn in all of the WDM runs suggests that even softenings of
$\epsilon/\epsilon_0$=1, while helping to suppress relaxation, cannot
eliminate it in realistic circumstances, where initial collapse will seldom
be symmetric. It also highlights the inherent difficulty in post-processing
halo catalogues to remove spurious haloes -- even in the most
conservative limit of $\epsilon/\epsilon_0$=1, evidence for numerical
artifacts persists.

In Figure~\ref{fig:differences}, we compare and contrast the
$\epsilon/\epsilon_0$=0.01 and 1 runs using difference maps (i.e. difference
in pixel by pixel values of projected density maps). This confirms the
impression from Figures~\ref{fig:cdm_haloes} and
~\ref{fig:wdm_haloes} -- that reducing the softening seeds the growth of
small-scale structure in the density field in the CDM runs, whereas this is
negligible in the WDM runs. This is the behaviour we would expect in the
presence of discreteness-driven relaxation of the kind that we have described.
Because there is power on all scales in
the CDM model, gravitational collapse proceeds early -- how early depends on
the scales that are resolved in the $N$-body simulation. In our runs
containing two collisionless components, the heavier particles seed local
gravitational perturbations that act to modify the velocity distributions
of particles; precisely how will depend on the local gravitationally induced
velocity field. The net effect, however, is to amplify density perturbations
and to cause regions to undergo premature gravitational collapse\footnote{We see
  a similar effect in WDM simulations in which a thermal velocity component
  is added explicitly to dark matter particles in the initial conditions
  to mimic the effect of free-streaming\citep[cf. \S2.1.2. of ][]{power.2013}}.
The lack of power on small scales in the WDM runs delays gravitational collapse
and so the effect of the localised gravitational and momentum perturbations
introduced by the heavier particles is delayed; the main differences arise
in the filaments, where the beading characteristic of spurious halo formation
is evident.

\begin{figure*}
  \centerline{
    \includegraphics[width=0.99\columnwidth]{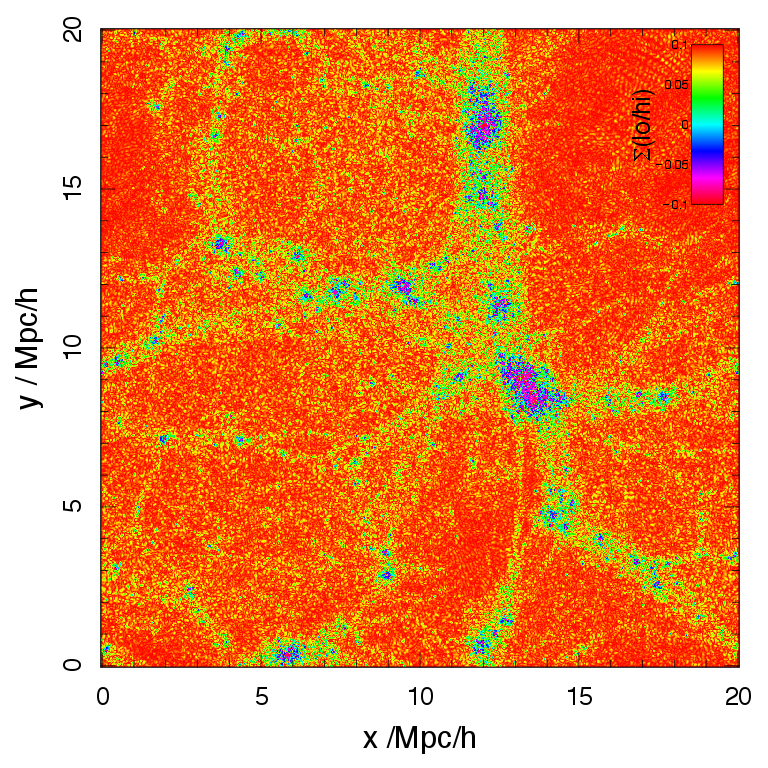}
    \includegraphics[width=0.99\columnwidth]{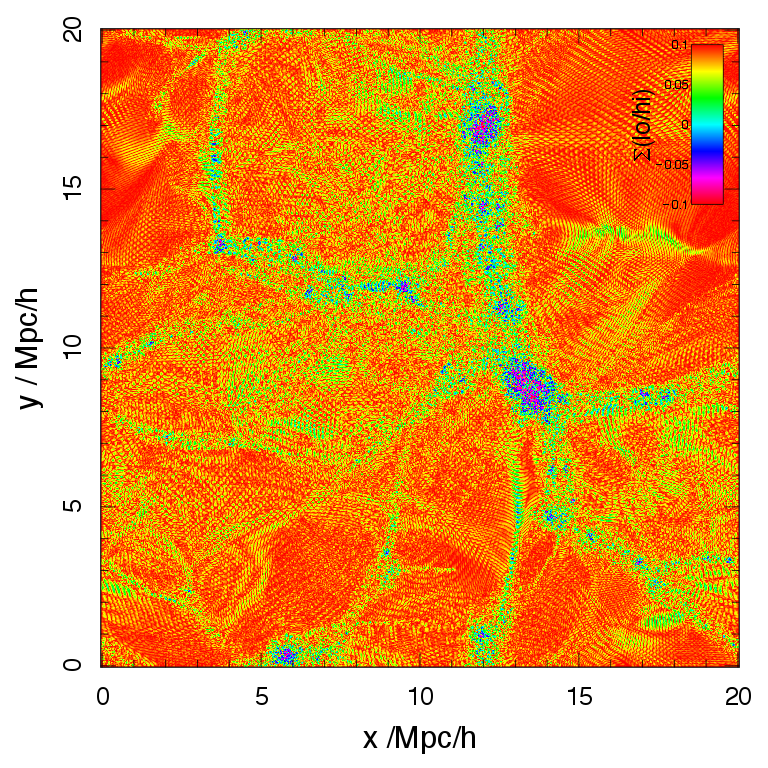}}
  \caption{{\bf Difference Maps: Delineating by Mass.} Here we show the
    difference in projected density between heavy and light particles in
    the $\epsilon/\epsilon_0$=0.01 CDM and WDM runs (left and right panels
    respectively). As before, we use the {\tt atan} mapping in the
      {\small R} {\tt magicaxis} package.}
  \label{fig:hilo-contrast}
\end{figure*}

In Figure~\ref{fig:hilo-contrast}, we investigate how the spatial distribution
of the two components (i.e. heavy and light particles) compare in the
$\epsilon/\epsilon_0$=0.01 CDM and WDM runs by looking at difference maps of
projected densities within pixels. This reveals that the
projected densities of the two components are similar within filaments and
FOF groups in both CDM and WDM runs. We expect the components to be well
mixed and to have similar projected densities within groups, where
gravitational collapse is well into its non-linear phase. Within filaments,
collapse is only mildly non-linear, and we might expect to see the components
clearly separated in projected density (similar to
Figure~\ref{fig:pancake_space}), as we see in the WDM run in the low density
regions. The pattern of gravitational collapse is plainly much more complex
in cosmological simulations than in the idealised plane-symmetric collapse
simulations of \S~\ref{ssec:pancake}, and so we might not expect such clear
separation in projected densities of the components; however, the
plane-symmetric collapse simulations with small softenings revealed that
perturbations at shell crossing seed momentum perturbations, which act to
smear out the separation between the initial planes of particles (and as the
two components collapse runs also showed, separations between the components),
and so we expect the components to be mixed within filaments too.
Interestingly, in the WDM run, it is striking how the filaments, and
consequently the areas where beading arising from spurious halo formation,
run orthogonally to the delineation in mass; this is much more difficult
to discern in the CDM run, where the long coherent threads of light and heavy
particles, so evident in the WDM run, are fragmented, with small-scale
structure present, and can be identified with the fog of lower-mass haloes
mentioned previously.

\begin{figure}
  \centering
  \includegraphics[width=0.99\columnwidth]{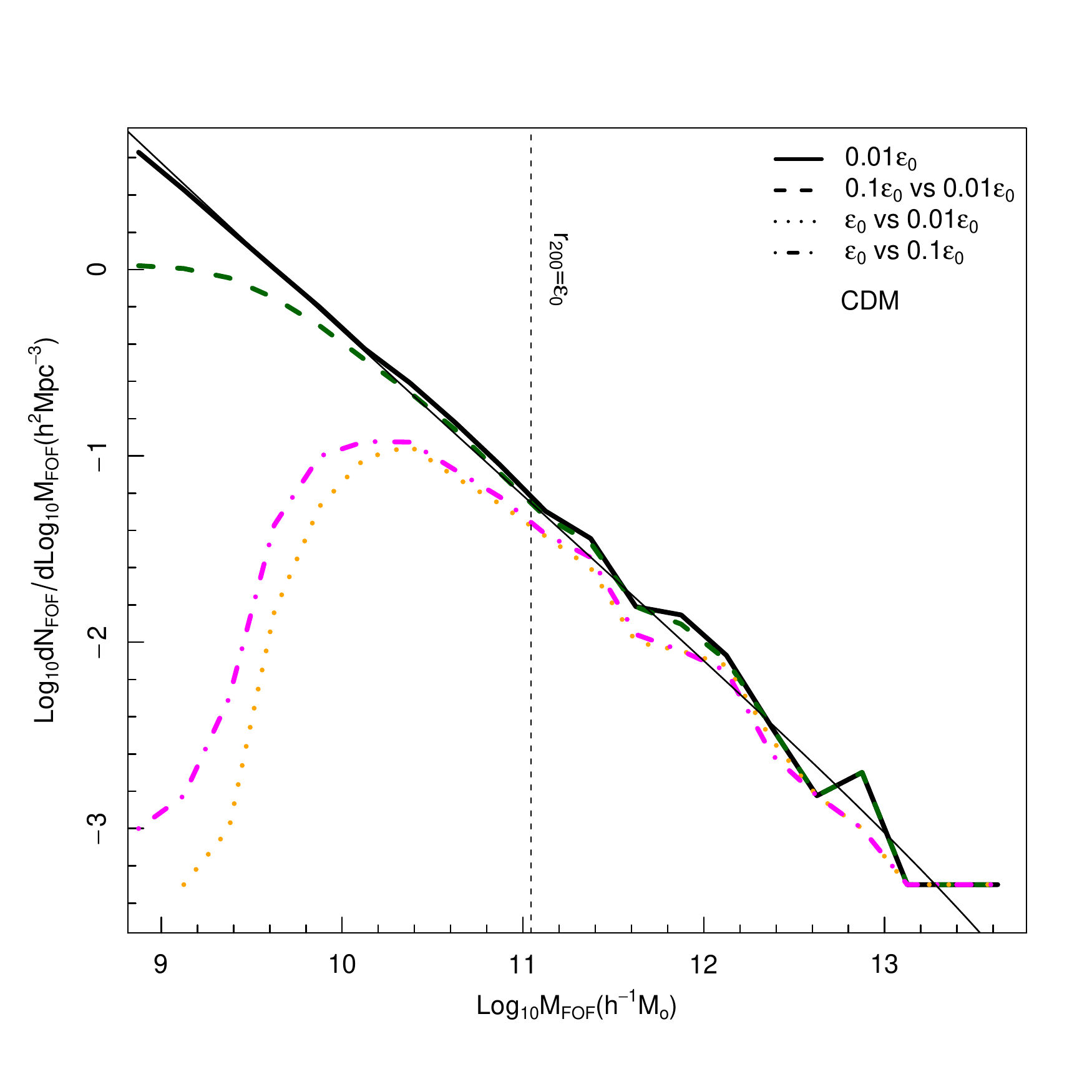}
  \centering
  \includegraphics[width=0.99\columnwidth]{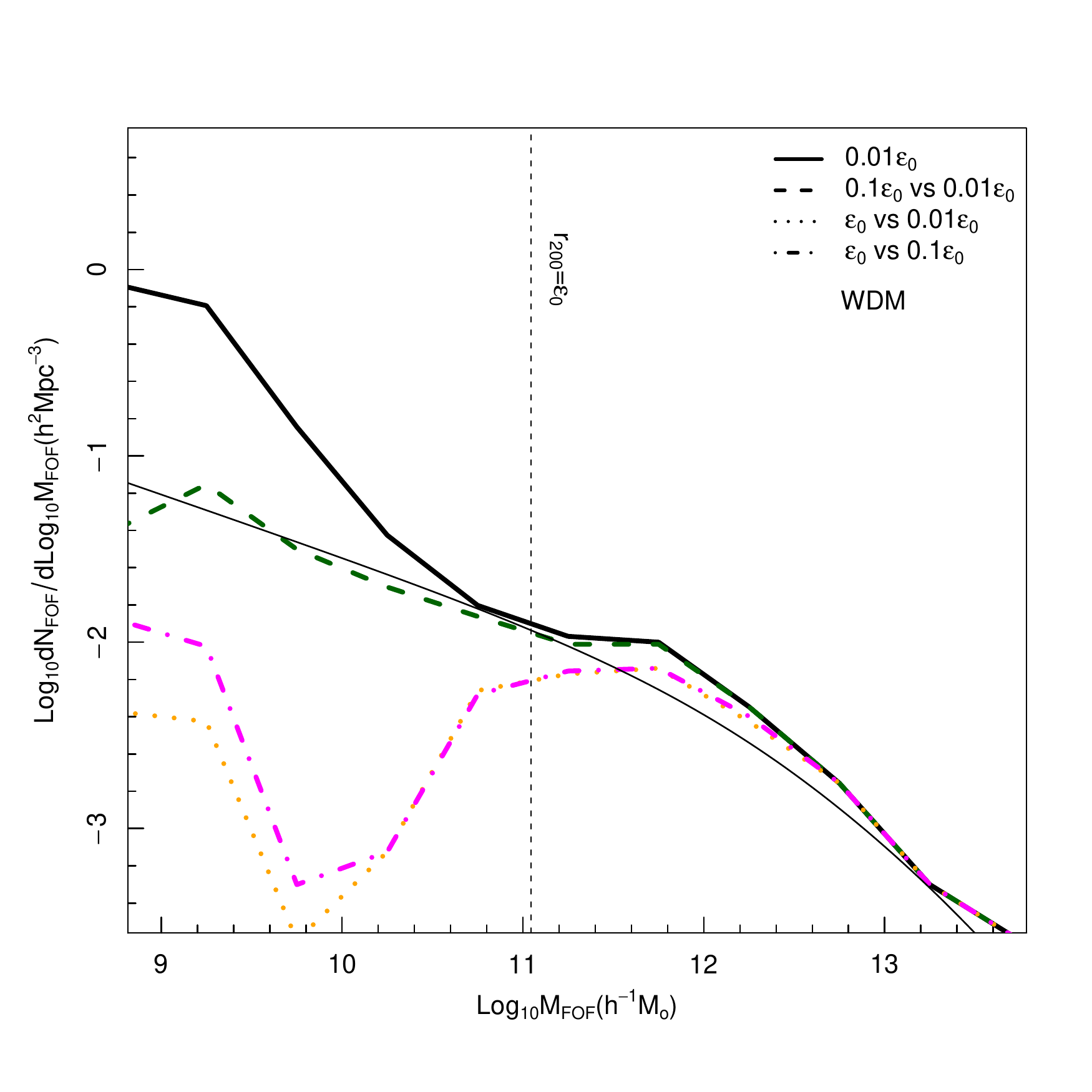}
  \caption{{\bf Matched FOF Group Mass Functions} Here we disentangle
    the contributions of spurious halos to the differential mass functions of
    FOF groups in the CDM runs (upper panel) and WDM runs (lower panel). The
    dashed, dotted, and dotted-dashed curves represent the number densities
    of halos that have been identified in the cross-matched FOF catalogues -
    0.1 to 0.01, 1 to 0.01, and 1 to 0.1 $\epsilon/\epsilon_0$; the heavy and
    light solid curves are the measured mass functions in the 0.01
    $\epsilon/\epsilon_0$ runs and the predictions from
    \citet{sheth.tormen.1999}. For reference, the vertical dashed line
    indicates the (approximate) minimum halo mass we might expect in
    the $\epsilon/\epsilon_0$=1 run, where the halo virial radius $r_{200}$
    is set to $\epsilon_0$. Recall that for $m_{\rm WDM}$=0.2 keV/$c^2$, the
    appropriate half-mode mass is
    $M^{\rm half}_{\rm WDM}\simeq 1.5 \times 10^{12} h^{-1} {\rm M_{\odot}}$.}
  \label{fig:matching_mfs}
\end{figure}

In Figures~\ref{fig:matching_mfs}, we investigate the consistency between
simulations of differing $\epsilon/\epsilon_0$ by cross matching FOF groups
in the CDM runs (upper panel) and WDM runs (lower panel) and directly
comparing particles (using {\small GADGET2} IDs), requiring that more than
90\% of particles to be in common. The heavy solid curve is the FOF group
mass function measured in the $\epsilon/\epsilon_0$=0.01 run, while the dashed,
dotted, and dotted-dashed curves correspond to the mass functions of halos
that are recovered in the (i) $\epsilon/\epsilon_0$=0.1 and 0.01, (ii) 0.01
and 1, and (iii) 0.1 and 1 runs respectively. The light dashed vertical
line indicates the limiting halo mass we might expect naively if the formation
of structure with $r_{200}$ smaller than $\epsilon$=$\epsilon_0$ is suppressed;
the limit corresponding to the 0.1$\epsilon_0$ lies off the plot. 
Recall that these results are for $m_{\rm WDM}$=0.2 keV/$c^2$, and so the
appropriate half-mode mass is
$M^{\rm half}_{\rm WDM}\simeq 1.5 \times 10^{12} h^{-1} {\rm M_{\odot}}$
(cf. Eq~\ref{eq:half_mode_mass}).

There are a number of interesting points worthy of note in these Figures. We
see in the CDM runs that the mass functions turn over at lower masses, as we
expect; in the $\epsilon/\epsilon_0$=0.1 to 1 and 0.01 to 1 runs, the
turn-over occurs at $\sim 1/10^{\rm th}$ the limit we naively expect to be
imposed by softening, equivalent to a few hundred particles, and there is
little difference between the two cases. In contrast, the turn-over in the
$\epsilon/\epsilon_0$=0.1 to 0.01 run becomes pronounced at $\sim 10$ larger
than the limit $r_{200}=0.1\,\epsilon_0$, equivalent to a few tens of particles.
This behaviour can be understood if one considers the WDM runs; here the
turn-over in the $\epsilon/\epsilon_0$=0.1 to 1 and 0.01 to 1 runs occurs very
close to the limit imposed by softening, although there is now an upturn at
lower masses. There is the hint of a turn-over in the $\epsilon/\epsilon_0$=0.1
to 0.01 run at low masses, in agreement with the CDM run, but we find that it
agrees well with the shape and amplitude of the \citet{sheth.tormen.1999}
mass function computed with {\small HMFcalc} \citep{murray.etal.2013}. Note
also that the turn-over in the  $\epsilon/\epsilon_0$=0.1 to 1 and 0.01 to 1
runs coincides with the upturn in the mass function that is attributed to
spurious halos.

The sharp turn-over at higher masses and the upturn at lower masses evident
in the $\epsilon/\epsilon_0$=0.1 to 1 and 0.01 to 1 runs reflects the cold
nature of collapse in the WDM model; the structures (typically linear)
contributing to the upturn collapse relatively late compared to their CDM
counterparts because of the reduced small scale power and are prevented from
collapsing further because of the influence of softening, but are still
sufficiently overdense to be linked together by the FOF criteria. This is not
so in the CDM case; here collapse proceeds at early times anisotropically,
and although softening helps to suppress smaller scale perturbations, they
are still sufficient to seed collapse on mass scales below the approximate
limit imposed by softening. There is no compelling physical reason why this
upturn in the WDM runs should be physical, and likely reflects the difficulty
the traditional $N$-body approach has in the limit where discreteness-driven
relaxation are important.

\begin{figure}
  \centering
  \includegraphics[width=8cm]{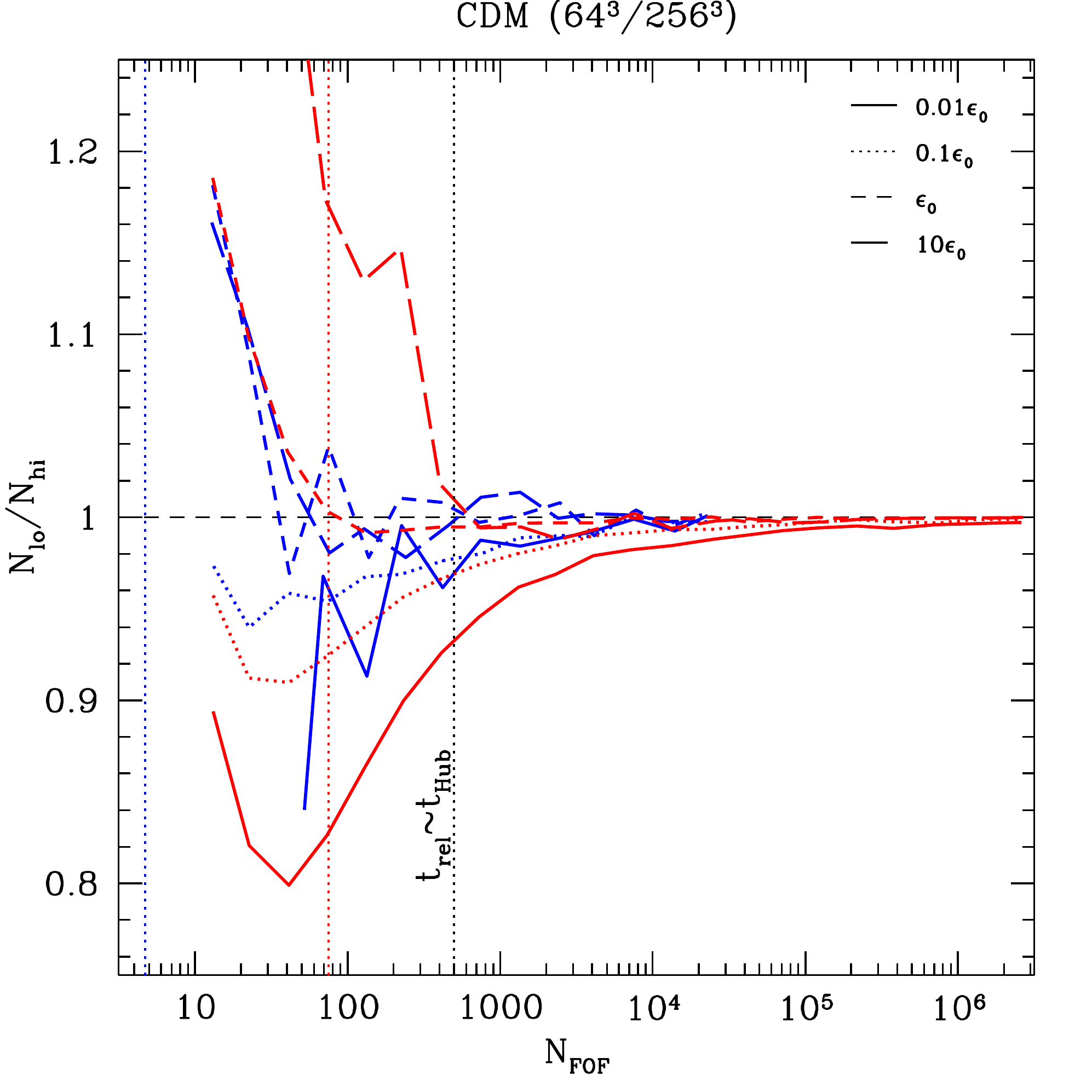}
  \centering
  \includegraphics[width=8cm]{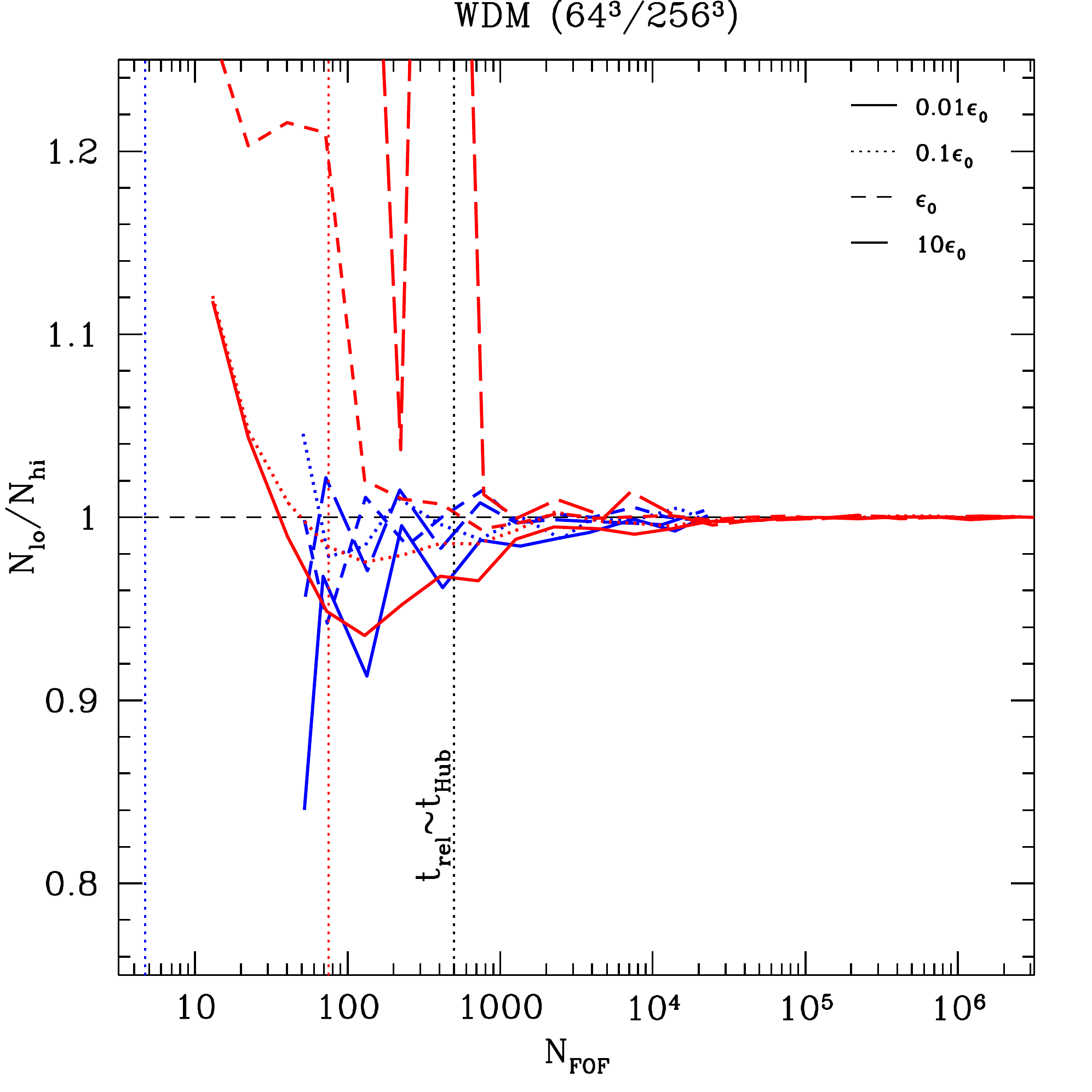}
  \caption{{\bf Evidence for Two-Body Relaxation in CDM and WDM Runs.} 
    Following \citet{binney.knebe.2002}, we use the ratio of the numbers of 
    low-to-high resolution particles (i.e. the numbers of heavy-to-light 
    particles) to measure the influence of two-body relaxation on the formation
    of haloes containing few particles. Solid, dotted, short- and long-dashed
    curves represent the $\epsilon/\epsilon_0$=0.01, 0.1, 1 and 10 CDM and
    WDM $1/4$ runs (upper and lower panels); curves 
    coloured red and blue correspond to runs with $2\times64^3$ and 
    $2\times256^3$ particles. The rightmost dotted vertical line indicates
    the size of FOF group ($N_{\rm FOF}$) below which the relaxation time is
    shorter than the Hubble time; the red and blue dotted vertical lines 
    indicate the value of $N_{\rm FOF}$ above which the \citet{wang.white.2007}
    criterion is satisfied.}
  \label{fig:f_ratio}
\end{figure}

\paragraph*{Halo Structure:} To complete our analysis, we follow BK02 and
consider the composition of the FOF groups by computing the ratio of the
numbers of low- and high-mass particles ($N_{\rm lo}$ and $N_{\rm hi}$
respectively). BK02 looked for the effects of mass segregation driven by
relaxation on the internal structure of haloes; in its absence, we would
expect uniformly mixed groups with $N_{\rm lo}/N_{\rm hi}$=1. In
Figure~\ref{fig:f_ratio} we compare the mean of the ratio
$N_{\rm lo}/N_{\rm hi}$ in each FOF group as a function of the 
total number of particles in the group ($N_{\rm FOF}$) for the CDM and WDM
runs (upper and lower panels), in bins of 0.25 dex, for softenings between
$\epsilon/\epsilon_0$=0.01 up to an extreme of 10; this is equivalent to 
Figure 4 of BK02. Vertical lines indicate:
\begin{itemize}
\item the number of particles corresponding to the \citet{wang.white.2007} 
  limiting mass for spurious haloes in WDM runs (blue is for the 
  $64^3$ runs, for direct comparison with BK02, while 
  red is for the $256^3$ runs).
\item the number of particles in a halo for which the relaxation time
  \begin{equation}
    t_{\rm relax} \sim \frac{0.1}{H_0} \frac{N_{\rm FOF}}{8 \log N_{\rm FOF}}
  \end{equation}
  is equal to the Hubble time ($t_{\rm Hub} \sim 1/H_0$) (for reference).
\end{itemize}
There are a number of interesting points worthy of note in
Figure~\ref{fig:f_ratio}. First, we find trends in our CDM runs
with $\epsilon/\epsilon_0 \le 0.1$ similar to those reported in
BK02, who interpreted them as evidence for two-body relaxation --
the ratio $N_{\rm lo}/N_{\rm hi}$ is order unity for FOF groups
containing $N_{\rm FOF} \gtrsim 1000$ particles, dips below 
unity for $N_{\rm FOF} \sim 100-1000$ particles, and rises sharply for 
$N_{\rm FOF} \lesssim 100$ particles. Second, softening alleviates the effects 
of two-body relaxation within groups -- compare the solid, dotted and dashed
curves; the smaller the softening, the larger the discrepancy between the
measured ratio and the ideal unity. Third, softening helps suppress two-body 
relaxation in larger groups, but it becomes ineffective once
$N_{\rm FOF}\lesssim 100$ particles; in larger groups, softening affects
structure in the sense that $N_{\rm lo}/N_{\rm hi} \lesssim 1$, i.e. higher
mass particles outnumber lower mass particles, which is what we would 
expect if mass segregation is occurring -- the higher mass particles contribute
a larger fraction of the halo mass, and so will be the dominant contribution 
to $N_{\rm FOF}$. Below 100 particles, the ratio 
$N_{\rm lo}/N_{\rm hi} \lesssim 1$ shows an upturn, which coincides with lower 
mass haloes on the outskirts of more massive haloes containing many of the low 
mass particles that have been displaced from the higher mass haloes through 
mass segregation. Fourth, the effect of too small a softening is less 
pronounced in WDM runs, but it is still evident. Interestingly, the
\citet{wang.white.2007} limiting mass tracks the upturn in the cross-matched
mass function that we see in the WDM runs in Figure~\ref{fig:matching_mfs}.

\subsection{Summary}
Taken together, these results demonstrate that discreteness-driven relaxation
affects the abundance of small-scale structure in all cosmological $N$-body
simulations. These effects are most readily apparent in WDM simulations, but
we see evidence in the CDM simulations for the same process occurring; however,
it is masked by the earlier onset of structure formation in CDM models on the
lowest mass scales, relative to the WDM model. Use of conservative gravitational
softenings $\epsilon=\bar{d}$ can help suppress the formation of unphysical
small-scale structure, at the expense of the internal structure of haloes, but
it cannot wholly eliminate it. This provides additional motivation for new
approaches
\citep[e.g.][]{hahn.2013a,angulo.2013,hahn.angulo.2015,hobbs.2015}.

\section{Conclusions}
\label{sec:summary}
We have explored the influence of discreteness-driven relaxation in
cosmological $N$-body simulations. Our motivation for doing this was to
understand what role such relaxation might play in the formation of spurious
haloes, most readily apparent in WDM simulations. We reasoned that the early
stages of gravitational collapse, which proceeds in an anisotropic fashion
as regions are sheared out into sheets and filaments \citep[cf.][]{kuhlman.etal.1996},
is susceptible to errors arising from discreteness effects because particles
move in a smooth mean-field gravitational potential modified by localised
perturbations, i.e. other particles. Encounters with these perturbations
modify the momentum distribution of particles and can seed the formation of 
unphysical structures. 

Using an idealised model for anisotropic collapse, the
plane-symmetric collapse test as set out in \citet{zeldovich.1970} and
\citet{shandarin.zeldovich.1989}, we demonstrated this effect, showing that
close encounters at shell crossing seeds the formation of unphysical clumpy
structures at later times, regularly spaced at intervals of the mean
inter-particle separation $\bar{d}$, which scales naturally as $m_p^{1/3}$. We
showed also that this effect depends on the magnitude of the kernel spline
gravitational softening scale $\epsilon$, and adopting
$\epsilon \sim \bar{d}=\epsilon_0$ can suppress the effect.
These effects were particularly striking when we
modelled the mass distribution using two collisionless components, in which
we varied the mass ratio between $1/1$ to $1/10$ and softenings between
$\epsilon/\epsilon_0$=(1,0.1,0.01), such that the larger the mass ratio
and the smaller the softening, the larger the perturbation at shell crossing.

We applied these insights to cosmological $N$-body simulations consisting of
two collionless components of differing fixed masses ($1/\sqrt{2}$ and $1/4$
respectively), designed to highlight the effect of discreteness-driven
relaxation, and found evidence for the effects that we observed in
plane-symmetric collapse simulations. These effects were most evident in the
WDM simulations, in the characteristic beads-on-a-string
structures within filaments and in the upturn of the FOF group mass function,
but they were also present in the CDM runs, most notably in the spatial
distribution of low-mass groups in low-density regions -- the so-called fog --
in runs with $\epsilon/\epsilon_0<1$, where we defined $\epsilon_0=\bar{d}$.
We found that choosing softenings $\epsilon/\epsilon_0$=1 could suppress the
effects of relaxation, but could not wholly eliminate them in realistic
circumstances. In the case of WDM runs, the dearth of small-scale power that
is present in CDM means that these effects will be most apparent
when collapse first becomes non-linear at relatively late times, in
larger-scale filaments; the momentum distribution of particles is unaffected
until then. In contrast, small-scale power drives collapse early in CDM runs
and so the momentum distribution of particles is modified at early times;
perturbations act to scatter momenta isotropically and this
is imprinted on the momentum distribution we would expect from linear
perturbation theory. In runs with small softenings, these momentum
perturbations are larger and give rise to the fog of low-mass groups
evident in Figure~\ref{fig:cdm_haloes}.

\medskip
In other words, the spurious halo problem is a generic one, affecting the CDM
model and its WDM and WDM-like counterparts alike. This is consistent with the
earlier work of \citet[][]{ludlow.porciani.2011}, who noted a
``missing progenitor'' problem in $N$-body simulations of the CDM model,
and with \citet{melott.etal.1997} and \citet{heitmann.etal.2005}, who found that
codes in which the gravitational force between particles is softened are unable 
to recover the analytical solution if the softening scale is smaller than the 
mean inter-particle separation, which \citet{melott.etal.1997} argued
represented collisionality. This is because the fundamental assumption
underlying $N$-body simulations -- that the system is collisionless and
integration of the equations of motion provides a solution to the Vlasov
equation -- ultimately falters because $N$ is finite, which implies that
discreteness effects will play a role in the evolution of the system.
This is leading to the development of new algorithms, such as approaches to
solve dark matter dynamics in phase space
\citep[cf.][]{hahn.2013a,sousbie.colombi.2015,hahn.angulo.2015}, and revisions
of the traditional $N$-body approach, such as spatially adaptive
\citep[e.g.][]{iannuzzi.dolag.2011,hobbs.2015} and anisotropic gravitational
softenings.

This has important implications for predictions of cosmological structure
formation and galaxy formation derived from traditional $N$-body
simulations. If discreteness-driven relaxation seeds
spurious halo formation, we might expect that the initial phases of halo
formation will tend to occur prematurely\footnote{This may be a naive
  expectation. For example, it may be more correct to say that it introduces
  a scatter between the predicted and numerically recovered anisotropic
  collapse times; depending on the details of collapse, these times may
  earlier or later than expected. It is also interesting to note that these
  may have consequences at late times. Previous studies
  \citep[e.g][]{vandenbosch.2002,giocoli.etal.2007,power.etal.2012},
  which compare the formation times of simulated halos and the
  predictions of extended Press-Schechter (EPS) theory, report that
  the formation times of simulated halos are systematically earlier than
  (EPS) predictions.}
impact halo structure
at later times, allowing for the influence of merger history. This is
because overly dense progenitors sink to the centre as they merge to form
more massive systems, and will artificially enhance central density
\citep[see also][]{binney.knebe.2002}. This is likely to 
lead to a modification of the well studied relationship between halo mass and
concentration \citep[cf.][]{nfw.1997,bullock.etal.2001,duffy.etal.2008,dutton.maccio.2014,ludlow.etal.2014,correa.etal.2015}
at lower masses -- the influence should be erased at higher masses through
the effects of repeated merging -- and suggests that the disparity reported
in the slope at lower masses of relations proposed by different authors
\citep[see discussion in][]{correa.etal.2015} could be numerically driven.
This also implies that low-mass haloes accreted later
could survive for longer because of their enhanced central densities,
possibly giving rise to many more substructures than we might otherwise
expect, although it is worth nothing that these are also the systems
whose central structure is most poorly resolved and so are subject to enhanced
numerically-driven tidal disruption.

In addition, the combination of earlier formation times and enhanced
survival during merging will influence the predictions of galaxy formation
models based on the semi-analytical approach coupled to $N$-body merger
trees \cite[cf.][]{baugh.2006} -- haloes can form stars earlier, and more of
these haloes containing stars will survive to merge with central galaxies.
This will have consequences for the predicted luminosity function and the
calibration of different modes of feedback, including active galactic nuclei
the most massive dark matter halos. Of course, the impact of this effect can
be gauged by means of Monte Carlo merger trees based on extended
Press-Schechter theory \citep[e.g.][]{parkinson.etal.2008}.
  
It also affects hydrodynamical simulations, where it is
standard practice in smoothed particle hydrodynamics (SPH) runs to use equal
numbers of gas and dark matter particles, which for typical cosmological
simulations leads to a mass ratio of $\Omega_b/(\Omega_0-\Omega_b) \simeq 1/6$
between them, where $\Omega_b$ is the baryon density parameter; when coupled
to small gravitational softenings, this can lead to both spurious heating
\citep[as has been know for some time; see, e.g.][]{steinmetz.white.1997}
as well as spurious growth (cf. \citealt{oleary.mcquinn.2012},
\citealt{angulo.2013a}).

We have outlined a number of important consequences of the effects of
discreteness-driven relaxation for predictions of cosmological structure
formation and galaxy formation. Broadbrush predictions will be unaffected --
for example, there are sound theoretical reasons to expect cuspy dark matter
halos \citep[e.g.][]{moore.1994,schulz.etal.2013} -- but the devil will be
in the detail -- for example, what is the true predicted abundance of CDM
dark matter subhalos around a galaxy like the Milky Way, and how should we
expect the internal structure of subhalos to be affected? These are strong
arguments for a more considered approach to the limitations of $N$-body
simulations, powerful a tool as they may be, and provides additional motivation
for the need of new approachs
\citep[e.g.][]{hahn.angulo.2015,hobbs.2015,sousbie.colombi.2015}.

\section*{Acknowledgments}
The authors are indebted to the referee whose careful reading of the paper
and their insightful suggestions have helped to improve it. CP, GFL and DO
acknowledge support of Australian Research Council (ARC) 
DP130100117. CP, ASGR, GFL and DO acknowledge support of ARC DP140100198. CP 
acknowledges support of ARC FT130100041. All simulations presented in this 
paper were carried out using computational resources on the Magnus
supercomputer at the Pawsey Supercomputing Centre through the National
Computational Infrastructure Merit Allocation Scheme. The research presented
in this paper is undertaken as part of the Survey Simulation Pipeline
(SSimPL; {\texttt{http://ssimpl.org/}).

\vspace{1cm} \bsp

\bibliographystyle{mn2e}

\appendix

\section{Influence of TreePM Algorithm}
We have demonstrated in \S~\ref{ssec:pancake} that particle discreteness
introduces gravitational perturbations when the gravitational softening
length is smaller than the mean inter-particle separation, which breaks
the symmetry inherent in the plane-symmetric collapse problem and
seeds the formation of spurious structures. This will be sensitive to
both the manner and the accuracy with which we calculate forces, the
errors in which can be asymmetric and which can in turn seed perturbations
that form spurious structures. This is especially true in the case of pure
Tree codes, where force errors are non-Cartesian. We have tested the
sensitivity of our results to the TreePM algorithm used in {\small GADGET2}
by running a subset of our simulations using a purely Tree calculation
(hereafter {\small NoPM}) as well as the default hybrid TreePM calculation
with increasing PM dimension, from $N_{\rm mesh}$=64 to 1024 in factors of 2
(hereafter {\small PM-$N_{\rm mesh}$}). Recall that we have used by default
the TreePM option with a PM dimension of 512.

\medskip

\paragraph*{Plane-Symmetric Collapse:} Here we assess how the $256^3$
version of the plane-symmetric collapse problem, with
$\epsilon/\epsilon_0$=0.1, is affected by influenced by our choice of
PM dimension by looking at the phase space structure at $z$=0 in
Figure~\ref{fig:phase_space_pm} and the corresponding projected spatial
structure in Figure~\ref{fig:spatial_structure_pm}. From top left to bottom
right, we show results for the {\small NoPM}, {\small PM-128}, {\small PM-256},
and {\small PM-512} runs; variations in CPU time per run were less than
$\sim 3\%$.

\smallskip

The clumping seeded at shell crossing that is evident in
Figure~\ref{fig:pancake_phase} is also apparent in
Figure~\ref{fig:phase_space_pm}, independent of our choice of PM dimension,
although the degree of clumping reduces as PM dimension increases.
A similar trend is evident in Figure~\ref{fig:spatial_structure_pm}. These
results suggest that the greater force asymmetries and inaccuracies
implicit in runs with coarser PM dimension amplify, rather than give rise to,
the discreteness-driven relaxation effects that we report.

\begin{figure*}
  \centerline{
    \includegraphics[width=0.99\columnwidth]{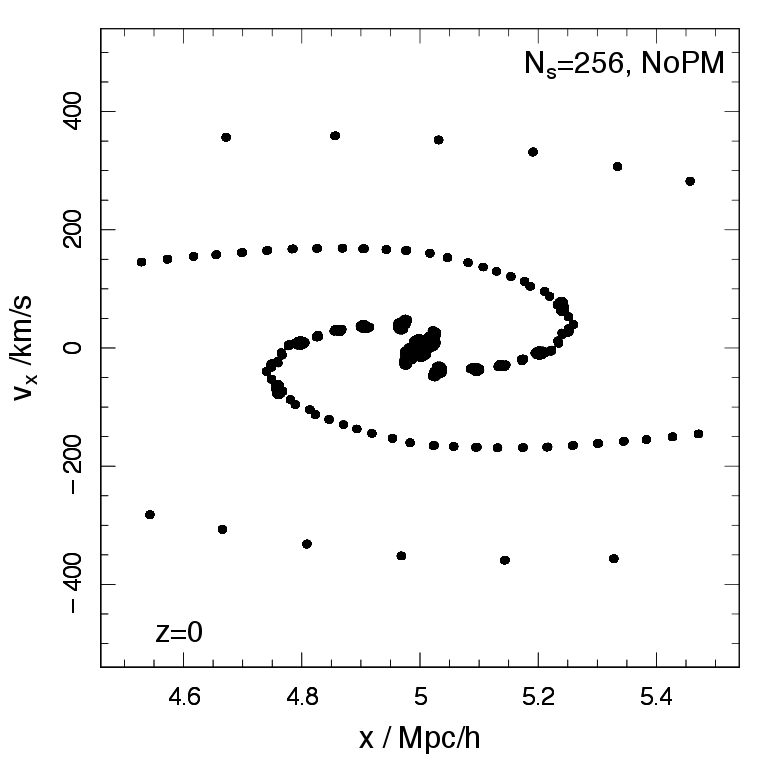}
    \includegraphics[width=0.99\columnwidth]{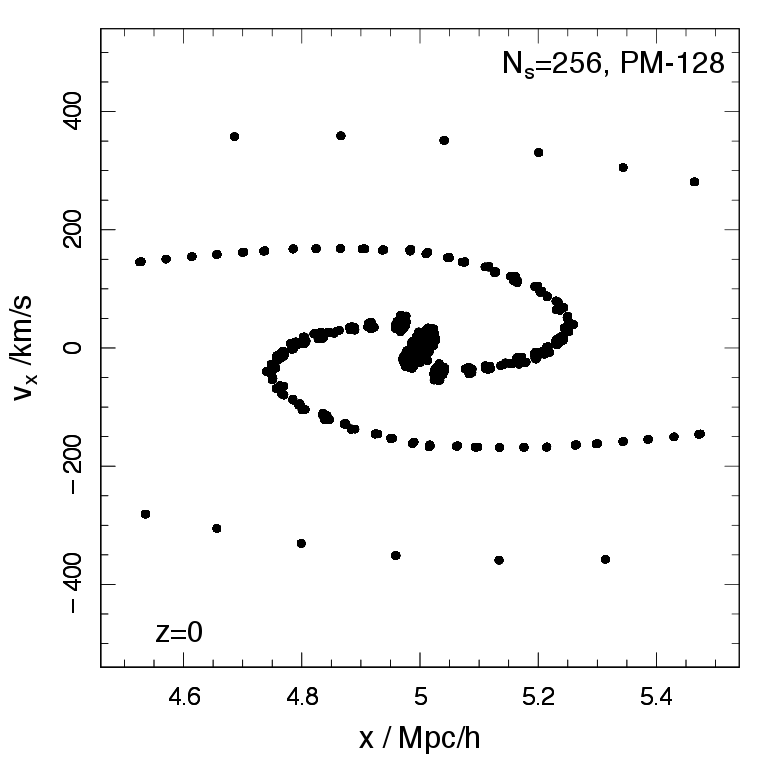}    }
  \centerline{
    \includegraphics[width=0.99\columnwidth]{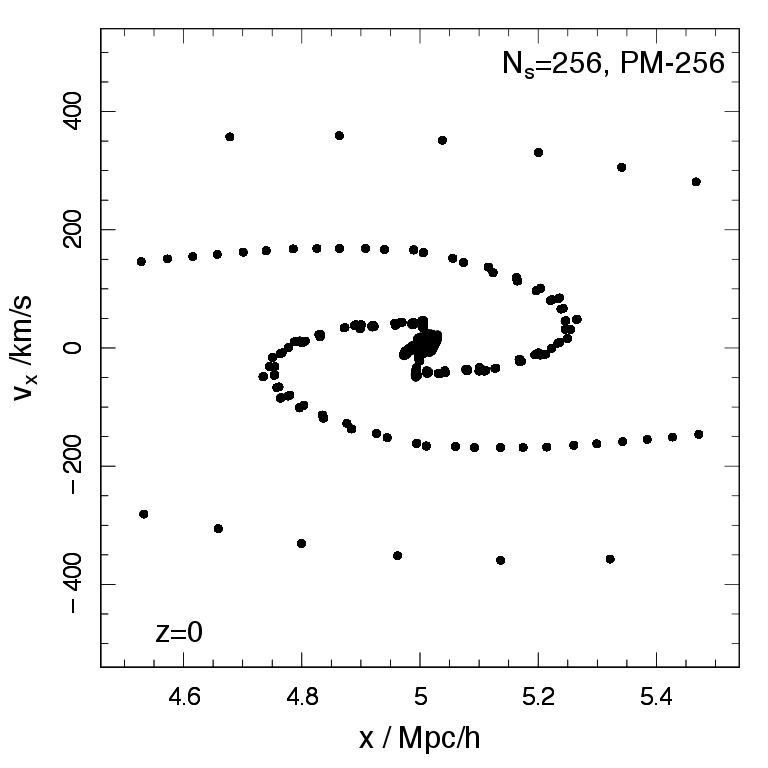}
    \includegraphics[width=0.99\columnwidth]{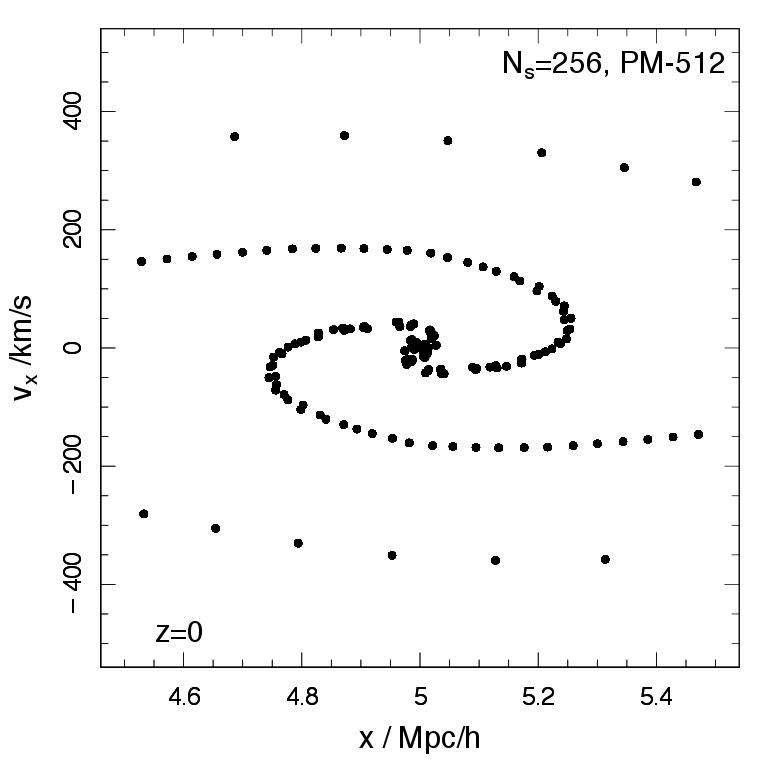}    }
  \caption{{\bf Plane-Symmetric Collapse: Phase Space Structure}. Here we show
    how the phase space structure in the $256^3$ run at approximately shell
    crossing ($z\simeq 4$) is affected by the PM dimension of the TreePM
    algorithm. As in Figure~\ref{fig:pancake_phase}, $v_x$ is the peculiar
    velocity along the $x$-direction and $x$ is the comoving position. The
    gravitational softening $\epsilon$=0.1$\epsilon_0$.}
  \label{fig:phase_space_pm}
\end{figure*}

\begin{figure*}
  \centerline{
    \includegraphics[width=0.99\columnwidth]{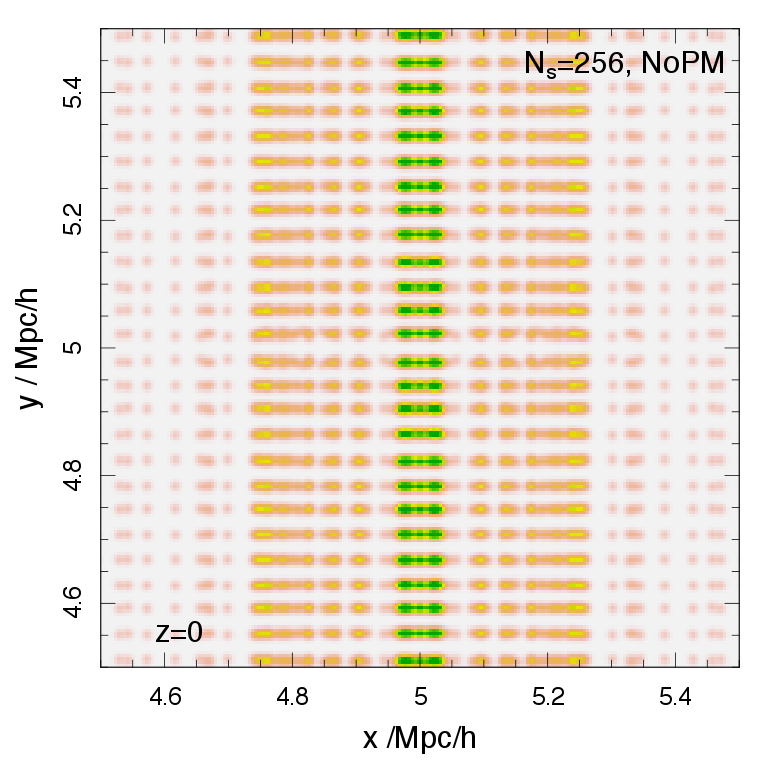}
    \includegraphics[width=0.99\columnwidth]{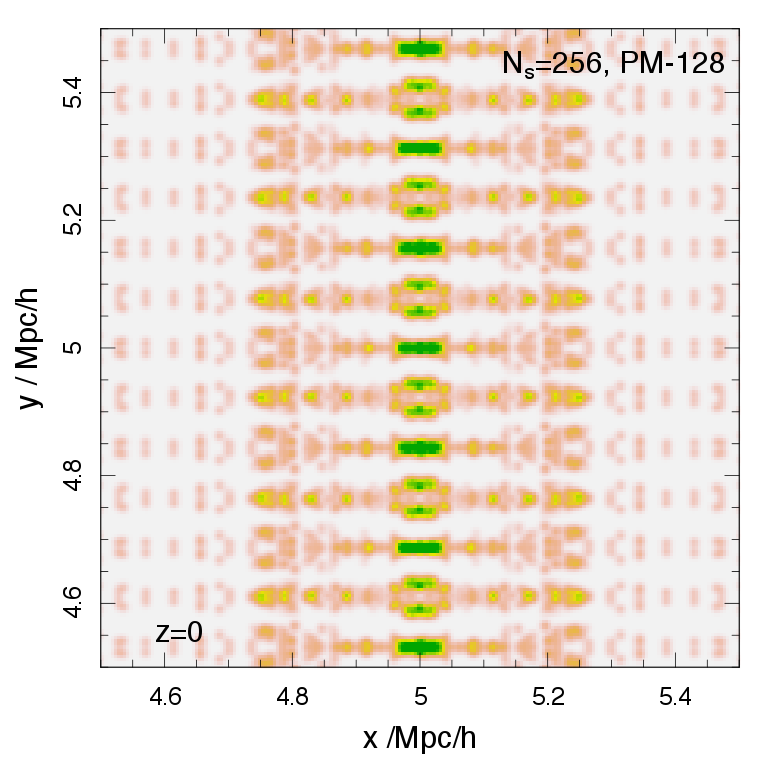}    }
  \centerline{
    \includegraphics[width=0.99\columnwidth]{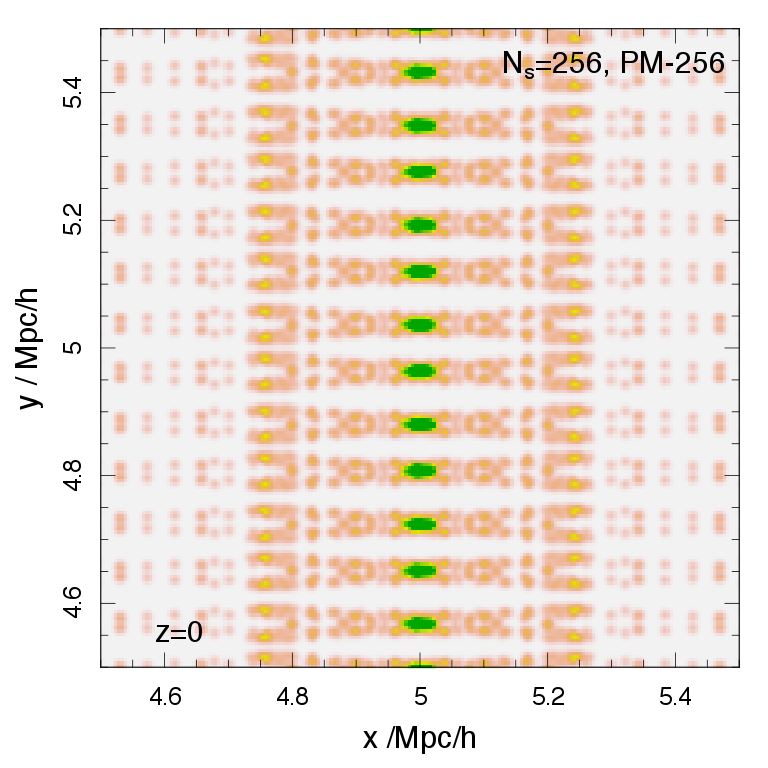}
    \includegraphics[width=0.99\columnwidth]{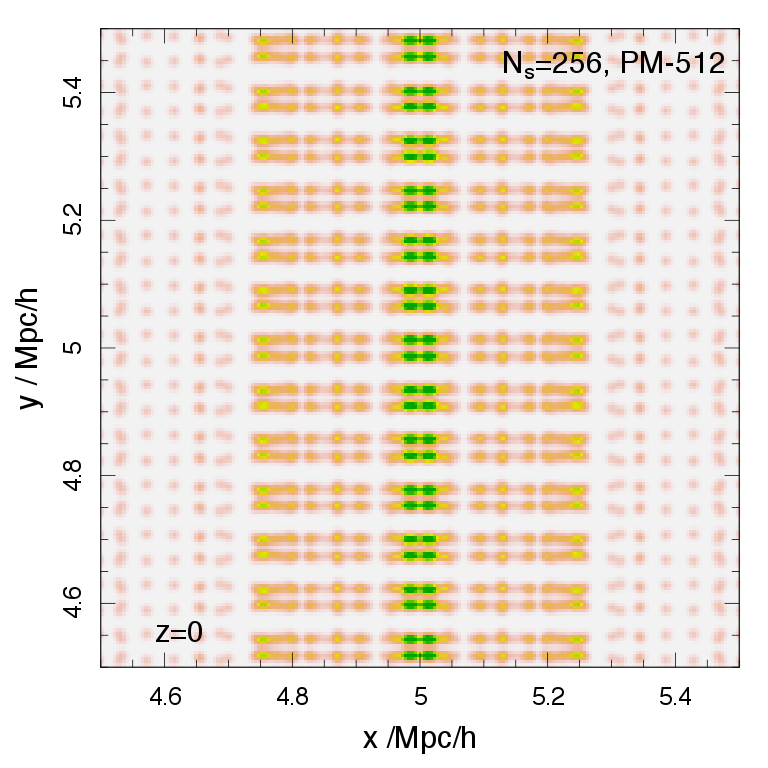}    }
  \caption{{\bf Plane-Symmetric Collapse: Spatial Structure.} Here we show
    how the projected spatial distribution (in the $x$-$y$ plane) at $z$=0
    in the $256^3$ run is affected by the PM dimension of the TreePM algorithm.
    Positions are in comoving coordinates, while the gravitational softening
    $\epsilon$=0.1$\epsilon_0$.}
  \label{fig:spatial_structure_pm}
\end{figure*}

\medskip

\paragraph*{Cosmological Simulations:} We now consider the CDM version of the
two collisionless component run with mass ratio of $1/4$ and
$\epsilon/\epsilon_0$=0.1, with a focus on the power spectrum of density
perturbations and on the abundance of FOF groups at $z$=0.

\smallskip
A few key diagnostics of the runs are presented in Table~\ref{tab:pm_tests}.
For the given problem size, the {\small PM-512} calculation is most
computationally efficient, while neglecting the PM component completely
results in a calculation that is half as efficient. There is little
variation the numbers of FOF groups with in excess of 1000 particles that
form, as we would expect; their formation is driven by the large scale
gravitational field, and although their evolution in detail may be affected
force inaccuracies, these will not be sufficient to significantly affect
which regions collapse and what masses they reach. In contrast, there is a
trend for the numbers of FOF groups with in excess of 10 and 100 particles
to decrease as finer TreePM calculations are used, with the PM cell dimension
matching the mean inter-particle separation as the transition point
(i.e. $N_{\rm mesh}$=256).

\begin{table}
\begin{center}
  \caption{\textbf{Sensitivity to TreePM Algorithm:} Here $N_{\rm}$ is the
    PM dimension along 1 dimension; $f_{\rm CPU}$ is the CPU time in units of
    the time taken for the {\small PM-1024} run; and $N_{\rm FOF}^{>x}$ is
    the number of FOF groups obtained with {\small SubFind} with a linking
    length of $0.2\,\bar{d}$ with in excess of $x=(10,100,1000)$ particles. For
    reference, all simulations were run on 16 processors; the {\small PM-1024}
    run, which took $\sim$186 CPU hrs.}
\vspace*{0.3 cm}

\begin{tabular}{lllccc}\hline
 & $N_{\rm PM}$  & $f_{\rm CPU}$ & $N_{\rm FOF}^{>10}$ & $N_{\rm FOF}^{>100}$ & $N_{\rm FOF}^{>1000}$ \\
\hline
{\small NoPM}    &  -   & 1.83 & 93282 & 8709 & 1223\\
{\small PM-64}   & 64   & 1.16 & 93702 & 8842 & 1251\\
{\small PM-128}  & 128  & 0.97 & 93806 & 8846 & 1235\\
{\small PM-256}  & 256  & 0.84 & 93508 & 8688 & 1220\\
{\small PM-512}  & 512  & 0.79 & 92384 & 8590 & 1229\\
{\small PM-1024} & 1024 & 1.   & 91632 & 8568 & 1232\\
\hline
\end{tabular}
\label{tab:pm_tests}
\end{center}
\end{table}

\smallskip
In Figure~\ref{fig:relative_pk} we show how the measured power spectra at
$z$=0 are affected by force errors by plotting the power spectra $P(k)$
between $k_{\rm min}=2\pi/L_{\rm box}\simeq 0.3 h\,{\rm Mpc}^{-1}$ and
$k_{\rm max}=k_{\rm Nyquist}=\pi\,N_{\rm mesh}/L_{\rm box}\simeq 3.1\times10^{-4} h\,{\rm Mpc}^{-1}$,
normalised to the power spectrum measured in the {\rm PM-1024} calculation,
$P(k)^{1024}$. Here $k_{\rm Ny}$ is the Nyquist frequency of the FFT mesh -
of dimension 512 - used to calculate the power spectrum, and we have used
the {\emph Cloud-in-Cell} mass assignment scheme when computing overdensities.
Dashed horizontal lines indicate a ratio of unity and $\rm 10\%$, while the
heavy solid, shorted dashed, dotted, dotted-dashed, and long dashed curves
correspond to the ratios obtained for the {\small NoPM}, {\small PM-64},
{\small PM-128}, {\small PM-256}, and {\small PM-512} runs respectively.
The {\small NoPM} run  shows the largest deviation, in excess of $10\%$ at
$k \simeq 10 h\,{\rm Mpc}^{-1}$, which is approximately the inverse mean
inter-particle separation of the simulation; otherwise the TreePM runs show
progressively smaller deviations with respect to the {\small PM-1024} run
from {\small PM-64} to {\small PM-128} to {\small PM-256}, at which point
there is little difference between the {\small PM-256} and {\small PM-512}
results. All of the runs show an excess at high wavenumber, approximately at
the scale of the inverse of the comoving softening.

\begin{figure}
  \centering
  \includegraphics[width=0.99\columnwidth]{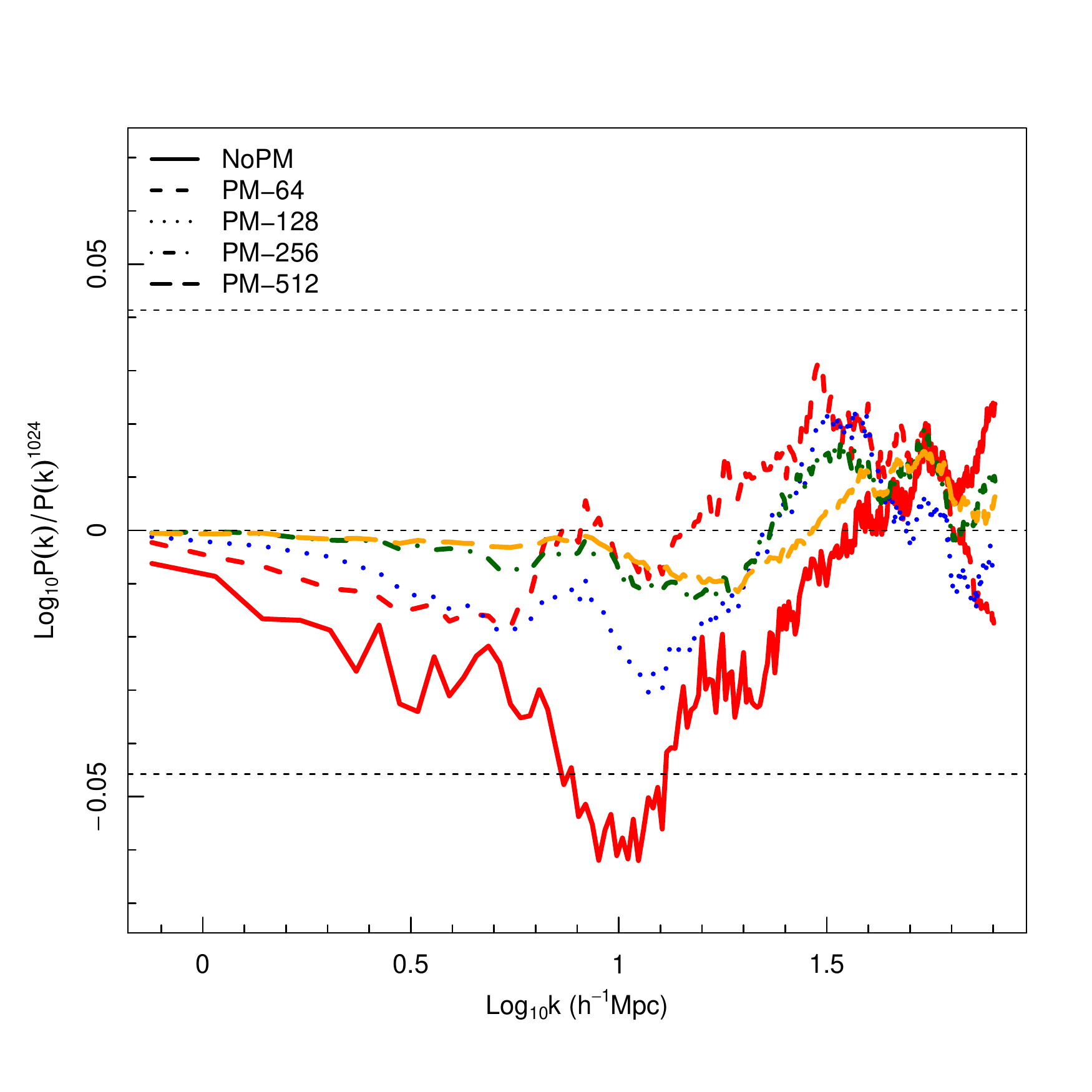}
  \caption{{\bf Impact on Power Spectrum:} Here we plot the power spectrum
    normalised to the power spectrum measured in the {\small PM-1024} run;
    solid, dashed, dotted, dotted-dashed, and long dashed curves correspond
    to the {\small NoPM}, {\small PM-64}, {\small PM-128}, {\small PM-256},
    and {\small PM-512} runs respectively. Dashed horizontal lines
    indicate unity and deviations of $\pm 10\%$.}
  \label{fig:relative_pk}
\end{figure}

\smallskip
In Figure~\ref{fig:relative_numbers}, we show how the FOF group mass
functions at $z$=0 are affected by the PM dimension by comparing the
abundance of halos cross matched with respect to the {\small PM-1024}
run to the abundance of all halos in the {\small PM-1024} run. The
mass functions are consistent down to a mass of
$\sim$ $10^{10.5} h^{-1} {\rm M}_{\odot}$, which
is equivalent to $\sim$$1000$ particles, before declining sharply
such that the deviation exceeds $\sim$$10\%$ at
$\sim$$10^{9.5} h^{-1} {\rm M}_{\odot}$, which is equivalent to $\sim$$100$
particles. These trends are in excellent agreement between runs, and,
as we have concluded above, this implies that force asymmetries and
inaccuracies amplify rather than give rise to the effects of
discreteness-driven relaxation.

\begin{figure}
  \centering
  \includegraphics[width=0.99\columnwidth]{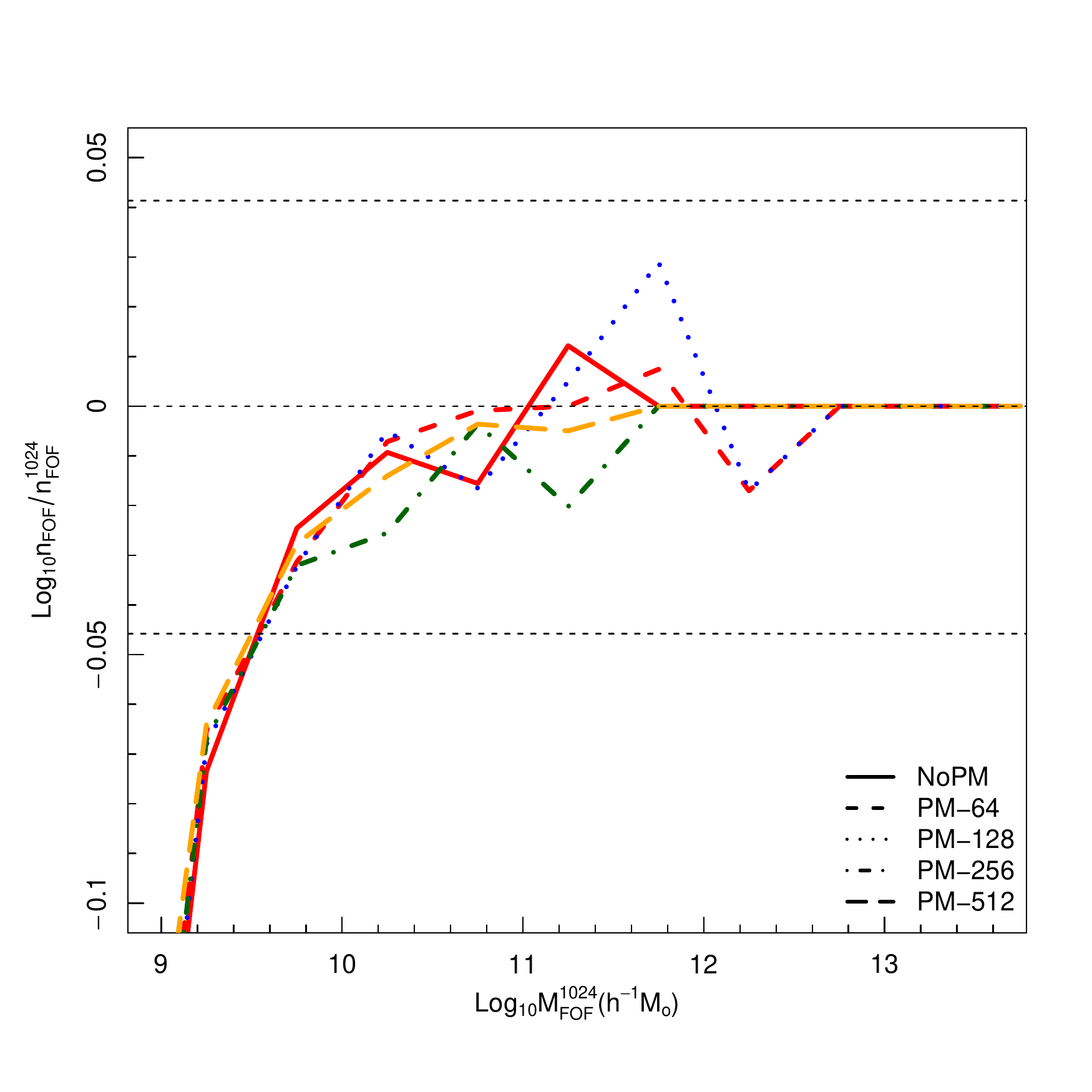}
  \caption{{\bf Impact on FOF Group Mass Function:} Here we plot the
    abundance of FOF groups (for a linking length of 0.2 $\bar{d}$)
    relative to that measured in the {\small PM-1024} run; solid, dashed,
    dotted, dotted-dashed, and long dashed curves correspond to the
    {\small NoPM}, {\small PM-64}, {\small PM-128}, {\small PM-256},
    and {\small PM-512} runs respectively. Dashed horizontal lines
    indicate unity and deviations of $\pm 10\%$.}
  \label{fig:relative_numbers}
\end{figure}

\label{lastpage}

\end{document}